\documentclass{article}
\usepackage{epsfig} 
\usepackage{amsmath}
\usepackage{amsfonts}
\usepackage{graphicx}

\usepackage{a4}

\usepackage{amsmath}

\begin{document}

\title{Electrically charged finite energy solutions of an $SO(5)$ and an $SU(3)$ Higgs-Chern-Simons--Yang-Mills-Higgs systems \\
in $3+1$ dimensions}
\author{{\large Francisco Navarro-L\'erida}$^{\ddagger}$
and {\large D. H. Tchrakian}$^{\star \dagger}$ 
\\ 
$^{\ddagger}${\small Departamento de F\'isica At\'omica, Molecular y Nuclear, Ciencias F\'isicas,}
\\
{\small Universidad Complutense de Madrid, E-28040 Madrid, Spain}
 \\ 
$^{\star}${\small School of Theoretical Physics, Dublin Institute for Advanced Studies,}\\
{\small 10 Burlington Road, Dublin 4, Ireland }\\
{$^\dagger$\small Department of Computer Science, Maynooth University, Maynooth, Ireland}}

\date{\today}
\newcommand{\dd}{\mbox{d}}
\newcommand{\tr}{\mbox{tr}}
\newcommand{\la}{\lambda}
\newcommand{\ka}{\kappa}
\newcommand{\f}{\phi}
\newcommand{\vf}{\varphi}
\newcommand{\F}{\Phi}
\newcommand{\al}{\alpha}
\newcommand{\ga}{\gamma}
\newcommand{\Ga}{\Gamma}
\newcommand{\de}{\delta}
\newcommand{\si}{\sigma}
\newcommand{\Si}{\Sigma}
\newcommand{\ta}{\theta}
\newcommand{\Ta}{\Theta}
\newcommand{\bnabla}{\mbox{\boldmath $\nabla$}}
\newcommand{\bomega}{\mbox{\boldmath $\omega$}}
\newcommand{\bOmega}{\mbox{\boldmath $\Omega$}}
\newcommand{\bsi}{\mbox{\boldmath $\sigma$}}
\newcommand{\bchi}{\mbox{\boldmath $\chi$}}
\newcommand{\bal}{\mbox{\boldmath $\alpha$}}
\newcommand{\bpsi}{\mbox{\boldmath $\psi$}}
\newcommand{\brho}{\mbox{\boldmath $\varrho$}}
\newcommand{\beps}{\mbox{\boldmath $\varepsilon$}}
\newcommand{\bxi}{\mbox{\boldmath $\xi$}}
\newcommand{\bbeta}{\mbox{\boldmath $\beta$}}
\newcommand{\ee}{\end{equation}}
\newcommand{\eea}{\end{eqnarray}}
\newcommand{\be}{\begin{equation}}
\newcommand{\bea}{\begin{eqnarray}}

\newcommand{\ii}{\mbox{i}}
\newcommand{\e}{\mbox{e}}
\newcommand{\pa}{\partial}
\newcommand{\Om}{\Omega}
\newcommand{\vep}{\varepsilon}
\newcommand{\bfph}{{\bf \phi}}
\newcommand{\lm}{\lambda}
\def\theequation{\arabic{equation}}
\renewcommand{\thefootnote}{\fnsymbol{footnote}}
\newcommand{\re}[1]{(\ref{#1})}
\newcommand{\R}{{\rm I \hspace{-0.52ex} R}}
\newcommand{\N}{{\sf N\hspace*{-1.0ex}\rule{0.15ex}%
{1.3ex}\hspace*{1.0ex}}}
\newcommand{\Q}{{\sf Q\hspace*{-1.1ex}\rule{0.15ex}%
{1.5ex}\hspace*{1.1ex}}}
\newcommand{\C}{{\sf C\hspace*{-0.9ex}\rule{0.15ex}%
{1.3ex}\hspace*{0.9ex}}}
\newcommand{\eins}{1\hspace{-0.56ex}{\rm I}}
\renewcommand{\thefootnote}{\arabic{footnote}}

\maketitle


\bigskip

\begin{abstract}

We study spherically symmetric finite energy solutions of two Higgs-Chern-Simons--Yang-Mills-Higgs (HCS-YMH) models in $3+1$ dimensions, one with gauge group $SO(5)$ and the other with 
$SU(3)$. The Chern-Simons (CS) densities are defined in terms of both the Yang-Mills (YM) and Higgs fields and the choice of the two gauge groups is made so they do not vanish.
The solutions of the $SO(5)$ model carry only electric charge and zero magnetic charge, while the solutions of the $SU(3)$ model are dyons carrying both electric
and magnetic charges like the Julia-Zee (JZ) dyon. Unlike the latter however, the electric charge in both models receives an important contribution from the CS
dynamics. We pay special attention to the relation between the energies and charges of these solutions.
In contrast with the electrically charged JZ dyon of the Yang-Mills-Higgs (YMH) system, whose mass is larger than that of the electrically neutral (magnetic monopole) solutions, the masses
of the electrically charged solutions of our HCS-YMH models can be smaller than their electrically neutral counterparts in some parts of the parameter space.
To establish this is the main task of this work, which is performed by constructing the HCS-YMH solutions numerically. In the case of the $SU(3)$ HCS-YMH, we have
considered the question of angular momentum, and it turns out that it vanishes.

\end{abstract}
\medskip
\medskip

\section{Introduction}
The main task of the present work is to establish that introducing CS dynamics to the YMH system can result in the lowering of the energy
of the electrically neutral solution, by giving it electric charge. We have tested this with two distinct models, one with gauge group $SO(5)$ and the other
$SU(3)$. The main difference between these two models is that, while the solutions of the $SU(3)$ model considered here carry magnetic charge, those of the $SO(5)$ model have zero
magnetic charge.

The two CS densities in $3+1$ dimensions employed here and in the preceding work~\cite{Navarro-Lerida:2013pua} are the first two in an infinite hierarchy, each
resulting from the descent \cite{Tchrakian:2010ar,Radu:2011zy} from a Chern-Pontryagin density in $2N$ ($N\ge 3$) dimensions. We refer to these as Higgs-CS (HCS) densities.
They extend the definition of the usual~\cite{Jackiw:1985,Deser:1982vy} CS densities to all odd and even dimensions, at the cost of importing a Higgs
field. In $2+1$ dimensions, it was found~\cite{Paul:1986ix,Hong:1990yh,Jackiw:1990aw} that the presence of the (usual) CS density
in a gauged Higgs system results in finite energy electrically charged solutions.
Here, the corresponding question is considered in $3+1$ dimensions.
It turns out that different choices of the HCS density employed, result in qualitatively quite different solutions. 

The $SO(5)$ HCS-YMH model considered here is that employed in a preceding work~\cite{Navarro-Lerida:2013pua}. In that preliminary work however, the energy of these solutions
increased with the electric charge, and the lowest energy solutions turned out to be those with vanishing charge. In this respect, they are qualitatively similar to
JZ solitons \cite{Julia:1975ff}. There the electrically neutral solutions had non-vanishing electric YM connection $A_0$, exhibiting dipole behaviour.

In the present paper we construct more general electrically charged solutions to this $SO(5)$ HCS-YMH model, some of which have lower energy than their
neutral counterparts, their energies decreasing with increasing charge. These qualitative features contrast with those of the
JZ dyons. As in \cite{Navarro-Lerida:2013pua}, there are also electrically neutral solutions exhibiting dipole behaviour,
namely supporting electrically neutral solutions with non-vanishing electric component of the YM connection $A_0$.
We have constructed three different families of solutions exhibiting these properties, which we refer to as Types I, II, and III.
Types I and II describe electrically charged solutions, while Type III solutions describe electrically neutral solutions with
non-vanishing electric component $A_0$ of the YM potential. None of these three types of solutions carry nonzero magnetic charge.

In addition to the $SO(5)$ HCS-YMH model, we have studied an $SU(3)$ HCS-YMH model. The main difference of the $SU(3)$ model is that
its solutions carry nonzero magnetic charge, at the same time supporting nonvanishing HCS terms. The resulting electrically charged
solitons are dyons which differ fundamentally from the JZ dyon. The feature of decreasing mass with increasing electrical charge,
observed for the solutions of the $SO(5)$ HCS-YMH model, persists also for the $SU(3)$ HCS-YMH model. In addition, we have considered the
question of angular momentum in the $SU(3)$ case.

The paper is organised as follows. In Section {\bf 2} we define the model, which is formally the same for both the $SO(5)$
and $SU(3)$ HCS-YMH models, except for the Higgs symmetry breaking potentials, which are stated there. Symmetry imposition on the
respective $SO(5)$ and $SU(3)$ HCS-YMH models is presented in Sections {\bf 3} and {\bf 4} respectively. In subsections of Sections
{\bf 3} and {\bf 4}, the numerical solutions are presented. Another subsection of Section {\bf 4} deals with the question of
angular momentum. Finally summary and discussion of our results are given in Section {\bf 5}.

\section{The models, equations, and charges} 
The full Lagrangian density is
\be
\label{CSYMH}
{\cal L}
= {\cal L}_{\rm YMH} +\ka_1\,\Omega_{\rm CS}^{(1)}+\ka_2\,\Omega_{\rm CS}^{(2)}\,,
\ee
with the two HCS densities $\Omega_{\rm CS}^{(1)}$ and $\Omega_{\rm CS}^{(2)}$ given by
\bea
\Omega_{\rm CS}^{(1)}&=&i\,\varepsilon^{\mu\nu\rho\si}\mbox{Tr}\,\F\,F_{\mu\nu}\,F_{\rho\si}\label{CS1} \,,\\
\Omega_{\rm CS}^{(2)}&=&i\,\varepsilon^{\mu\nu\rho\si}\mbox{Tr}\bigg[
\F\left(\eta^2\,F_{\mu\nu}F_{\rho\si}+\frac29\,\F^2\,F_{\mu\nu}F_{\rho\si}+\frac19\,F_{\mu\nu}\F^2F_{\rho\si}\right)
\nonumber\\
&&\qquad\qquad\qquad\qquad
-\frac29\left(\F D_{\mu}\F D_{\nu}\F-D_{\mu}\F\F D_{\nu}\F+D_{\mu}\F D_{\nu}\F\F\right)F_{\rho\si}\bigg]\, , \label{CS2}
\eea
where $\epsilon^{\mu\nu\rho\si}$ is the Levi-Civita tensor in Minkowski spacetime. We do not describe the provenance of the HCS
terms Eqs.~\re{CS1} and \re{CS2}, since this was given in detail in Appendix {\bf A} of Ref.~\cite{Navarro-Lerida:2013pua}. The role the Higgs scalar
plays here is somewhat akin to that of the axion~\cite{Peccei:1977ur,Peccei:1977hh}.

The YMH Lagrangian density is~\footnote{Since we aspire here to present a $3+1$ dimensional analogue of the the $2+1$ dimensional Chern-Simons-Higgs
vortices~\cite{Hong:1990yh,Jackiw:1990aw}, it may be relevant to inquire whether we could likewise omit the Yang-Mills term in Eq.~\re{CSYMH1}. This in principle
is possible since the system excluding the Yang-Mills term is consistent with the Derrick scaling requirement in the corresponding static Hamiltonian after
solving for $A_0$ using the Gauss-Law equation. However in the non-Abelian system at hand, $A_0$ cannot be solved for in closed form, rendering such an
approach impractical.}
\be
\label{CSYMH1}
{\cal L}_{\rm YMH}=\mbox{Tr}\left[\frac14F_{\mu\nu}^2-\frac12\,D_{\mu}\F^2-\frac{\la}{2}\,V[\eta^2,\F^2]\right]\,,
\ee
where $D_\mu = \partial_\mu + [A_\mu, \cdot]$.

Here, $V[\eta^2,\F^2]$ is the positive definite Higgs selfinteraction potential, with $\la$ its coupling constant,
and $\eta$ denoting the vacuum expectation value of the Higgs field. $\ka_1$ and $\ka_2$ are the coupling strengths of the HCS densities.

The equations of motion resulting from the variations of the Lagrangian with respect to the YM potential and the Higgs field are
\bea
D_{\mu}F^{\mu\nu}+[\F,D^{\nu}\F]&=&2\,i\,\ka_1\,\vep^{\mu\nu\rho\si}\,\{F_{\rho\si},D_{\mu}\F\}\,,\label{varYM}\\
D_{\mu}D^{\mu}\F-\la\{\F,(\F^2+\eta^2\eins)\}&=&i\,\ka_1\,\vep^{\mu\nu\rho\si}\,F_{\mu\nu}\,F_{\rho\si}\,,\label{varH}
\eea
respectively. $\{\ , \ \}$ denotes the anticommutator. These equations, Eqs.~\re{varYM} and \re{varH}, are written only for the Lagrangian with
$\kappa_2=0$ in Eq.~\re{CSYMH}. This is because the expressions for the right-hand sides of the corresponding equations for $\kappa_2\neq 0$ are very
cumbersome.

There are two types of symmetry breaking potentials consistent with the requirement of finite energy, which we list here for completeness
\bea
V_1&=&\left(\eta^2+a_1\,\mbox{Tr}\F^2\right)^2 \, , \label{V_1}\\
V_2&=&\frac14\,\mbox{Tr}\left(\eta^2\eins+a_2\F^2\right)^2\, ,\label{V_2}
\eea
where the values of $a_1$ and $a_2$ will be chosen according to our convenience when imposing symmetries. As it turns out, we will
concentrate mainly on $\la=0$ solutions since the presence of the HCS terms, Eqs.~\re{CS1}-\re{CS2}, is sufficient to support nontrivial field
configurations outside of $SU(2)$. When we do employ a potential for the purpose of checking that our conclusions are not altered by
the presence of one, then our choice is Eq.~\re{V_1} for both the $SO(5)$ and $SU(3)$ models.

The definition of the magnetic monopole charge is
\be
\label{tH-P}
\mu=-\frac{1}{4\pi}\,\vep_{ijk} \int_{S^\infty}\mbox{Tr}\,\F\,F_{ij}\, dS_k\,,
\ee
which presents a lower bound on the energy integral, and
the definition of the electric charge is
\be
\label{elec}
Q=-\frac{1}{4\pi}\,\int_{S^\infty}\mbox{Tr}\,\F\,F_{i0}\, dS_i\,.
\ee
The definitions Eqs.~\re{tH-P} and \re{elec} are valid, both when $\ka_1=\ka_2=0$, and when $\ka_1\neq 0$ and/or $\ka_2\neq 0$.

\section{Solutions of the $SO(5)$ Higgs-Chern-Simons--Yang-Mills-Higgs model}
This Section consists of two Subsections. In Subsection {\bf 3.1}, spherical symmetry is imposed and the boundary values of the solutions sought are
stated. The numerical construction of the solutions\footnote{We have employed a
collocation method for boundary-value ordinary differential equations, equipped with an adaptive mesh selection procedure
\cite{COLSYS}. A compactified radial coordinate $x=r/(1+r)$ has been used. Typical mesh sizes include $10^3-10^4$ points.
The solutions have a relative accuracy of $10^{-8}$.} is presented in Subsection {\bf 3.2}.

\subsection{Imposition of symmetry and boundary values}
To proceed to the imposition of symmetry, we note that the fields take their values in the $4\times 4$ chiral Dirac
representation of $SO(6)$
\bea
A_{\mu}&=&A_{\mu}^{\al\beta}\,\Si_{\al\beta}\ ,\quad \al=i,4,5;\ (i=1,2,3) \,, \label{so5}\\
\F&=&\psi^{\al\beta}\,\Si_{\al\beta}+\f^{\al}\,\Si_{\al 6}\,,\label{so6}
\eea
where $(\Si_{\al\beta},\Si_{\al 6})$ are the $4\times 4$ chiral representation matrices of $SO(6)$~\footnote{
The chiral Dirac representation matrices $\Sigma_{\mu\nu}=(\Sigma_{\al\beta},\Sigma_{\al 6})$ used here are defined as
$
\label{sigma1}
\Sigma_{\mu\nu}=-\frac14\Sigma_{[\mu}\,\tilde\Sigma_{\nu]}\,,
$
in terms of the spin matrices $\Sigma_{i}=-\tilde\Sigma_{i}=i\gamma_i\ ,\
\Sigma_{4}=-\tilde\Sigma_{4}=i\gamma_4\ ,\ \Sigma_{5}=-\tilde\Sigma_{5}=i\gamma_5\ ,\
\Sigma_{6}=+\tilde\Sigma_{6}=\eins$, where ($\gamma_{i},\ga_4,\ga_5$), $i=1,2,3$ are the usual Dirac gamma
matrices in four dimensions.}.

It is convenient to express our Ansatz using the index notation
$\al=i,M\ ,\quad i=1,2,3\ ,\quad M=4,5\,.$
With this notation, the static spherically symmetric Ansatz for the Higgs field $\F$, Eq.~\re{so6}, and the YM connection $A_{\mu}=(A_0,A_i)$ , Eq.~\re{so5}, are
\bea
\F&=&2\eta\left[\left(\f^M\,\Si_{M6}+\f^6\,\hat x_j\,\Si_{j6}\right)
-\left((\vep\psi)^M\,\hat x_j\,\Sigma_{jM}+\psi^{6}\,\Sigma_{45}\right)\right]\, ,\label{higgs}\\
A_0&=&-(\vep\chi)^M\,\hat x_j\,\Sigma_{jM}-
\chi^{6}\,\Sigma_{45}\, ,\label{a0p}\\
A_i&=&\left(\frac{\xi^{6}+1}{r}\right)\Sigma_{ij}\hat x_j+
\left[\left(\frac{\xi^M}{r}\right)\left(\delta_{ij}-\hat x_i\hat x_j\right)+
(\vep A_r)^M\,\hat x_i\hat x_j\right]\Sigma_{jM} +A_r^{6}\,
\hat x_i\,\Sigma_{45} \,, \label{aip}
\eea
in which the sum over indices $M,N=4,5$ runs over two values such that
we can label the functions $(\f^M,\f^{6})\equiv\vec\f$,
$(\chi^M,\chi^{6})\equiv\vec\chi$, $(\xi^M,\xi^{6})\equiv\vec\xi$, $(\psi^M,\psi^{6})\equiv\vec\psi$ and $(A_r^M,A_r^{6})\equiv\vec A_r$,
$i.e.$, in terms of five isotriplets $\vec\f$, $\vec\chi$, $\vec\xi$, $\vec\psi$, and $\vec A_r$, all depending on
the $3$ dimensional spacelike radial variable $r$. $\vep$ being the two
dimensional Levi-Civita symbol.

The full one dimensional subsystems are presented in Appendix {\bf A.1}.
It immediately follows from Eqs.~\re{higgs} and \re {fij} that the magnetic monopole charge Eq.~\re{tH-P}
vanishes.

The important quantity for us here is the global electric charge, Eq.~\re{elec}, which does not vanish. A straightforward calculation yields
the electric field $E_i$
\be
E_i=\mbox{Tr}\,\F\,F_{i0}=-2\eta\,\vec\psi\cdot D_r\vec\chi \, , \label{elec_field}
\ee
resulting in the electric charge
\be
\label{elec1}
Q=\frac{-1}{4^2\pi}\,\int_{S^\infty}\mbox{Tr}\,\F\,F_{i0}\, dS_i= \frac12\eta\,\left[r^2\,\vec\psi\cdot D_r\vec\chi\right]_{r=\infty}\,.
\ee

For both potentials Eqs.~\re{V_1} and \re{V_2}, the finiteness of the energy requires that 
\be
\label{Higgsasym}
\lim_{r\to\infty}\left(|\vec\f|^2+|\vec\psi|^2\right)=1 \, ,
\ee
so we can introduce an asymptotic angle $\gamma$ such that
\bea
\lim_{r\to\infty}\,|\vec\f|^2&=&\cos^2\ga\label{ga1} \, ,\\
\lim_{r\to\infty}\,|\vec\psi|^2&=&\sin^2\ga\label{ga2}\,.
\eea

The $SO(3)$ freedom in this Ansatz results in an invariance at the fixed point of the $2$-sphere, due to
which only two of the components of each of the five triplets $(\vec{A}_r,\vec\xi,\vec\chi,\vec\psi,\vec\f)$ are independent functions. We thus
end up with $10$ equations of motion for the functions of $r$,
\be
\label{mult}
\vec{A}_r=(\tilde a_r,0,a_r)\ ,\ \vec\xi=(\tilde  w,0,w)\ ,\ \vec\chi=(\tilde V,0,V)\ ,\ \vec\psi=(\tilde h,0,h)\ ,\ \vec\f=(\tilde g,0,g) \, .
\ee 
The equations of motion arising from the variation of $\vec{A}_r$ result in a pair of constraint equations, since there is no non-trivial
curvature pertaining to this connection.

We will study three types of solutions, for which these constraint equations are identically satisfied, such that
$\vec{A}_r=\vec{0}$ effectively. These finite energy solutions may have a non-vanishing electric charge and zero magnetic charge. It is straightforward to check that
the magnetic charge density in Eq.~\re{tH-P} vanishes identically for the field configuration parametrised by our spherically symmetric Ansatz, Eqs.~\re{higgs}, \re{a0p}, and \re{aip}.

These three types of zero magnetic charge solutions are described by the following functions
\bea
&&\vec\xi=(0,0,w)\ ,\ \vec\chi=(\tilde V,0,0)\ ,\ \vec\psi=(\tilde h,0,0)\ ,\ \vec\f=(\tilde g,0,0)\, ,\label{type1}\\
&&\vec\xi=(0,0,w)\ ,\ \vec\chi=(0,0,V)\ ,\ \vec\psi=(0,0,h)\ ,\ \vec\f=(0,0,g)\, ,\label{type2}\\
&&\vec\xi=(0,0,w)\ ,\ \vec\chi=(0,0,V)\ ,\ \vec\psi=(\tilde h,0,0)\ ,\ \vec\f=(0,0,g)\, ,\label{type3}
\eea
to which we refer as Types I, II, and III, respectively.
Such solutions exist for models with either of the Higgs potentials, Eqs.~\re{V_1} and \re{V_2}.

\subsection{Types I, II, and III: Numerical results}
We have not been able to generate numerically excited solutions when all the components in the multiplets Eq.~(\ref{mult}) are present. Only solutions for the restricted cases Eqs.~\re{type1}-\re{type3} could be found\footnote{We have set $\eta=1/2$ is our numerical schemes. This choice gives rise to a unit energy for type I solutions with $\la=0$, $\ka_1=0$, $\ka_2=0$, and $\ga=\pi/2$.}. 

\subsubsection{Type I solutions}

These solutions are characterized by $\tilde w = 0$, $V=0$, $h=0$, and $g=0$. The expansions at the origin are

\bea
&&w = -1 + w_2 x^2 + 2 w_2 x^3 + O(x^4) \, , \\
&&\tilde V = \tilde V_1 x + \tilde V_1 x^2 + O(x^3) \, , \\
&&\tilde g = \tilde g_0 + O(x^2) \, , \\
&&\tilde h = \tilde h_1 x + \tilde h_1 x^2+ O(x^3) \, , \label{exp_or_type_I}
\eea
where $x= r/(1+r)$. The asymptotic values of the functions are
\bea
&&w = 0 \ , \\
&&\tilde V = \tilde V_0 \, , \\
&&\tilde g = \cos\gamma \, , \\
&&\tilde h = \sin\gamma \, , \label{exp_inf_type_I}
\eea
where $\tilde V_0$ and $\gamma$ are free. $\tilde V_0$ controls the contribution to the electric charge Eq.~(\ref{elec}) of JZ type, while $\gamma$ gives rise to another contribution to the electric charge, once the HCS terms are present. Our parameters are: $\lambda$, $\kappa_1$, $\kappa_2$, $\tilde V_0$, and $\gamma$.

The effect of the JZ parameter  $\tilde V_0$ is exhibited in Fig.~1. For fixed $\ga$ and $\ka_1=0$ and $\ka_2=0$ , when varying $\tilde V_0$ the electric charge $Q$ changes. In this case an increase in $|Q|$ makes the energy of the solutions $E$ increase. This is the behaviour one would expect. In fact, for vanishing $\la$ the theory may be rescaled and the relation between $E$ and $Q$ becomes independent of $\ga$ (they both rescale with $\sin\ga$).

\begin{figure}[ht]
\begin{center}
\includegraphics[width=0.7\textwidth]{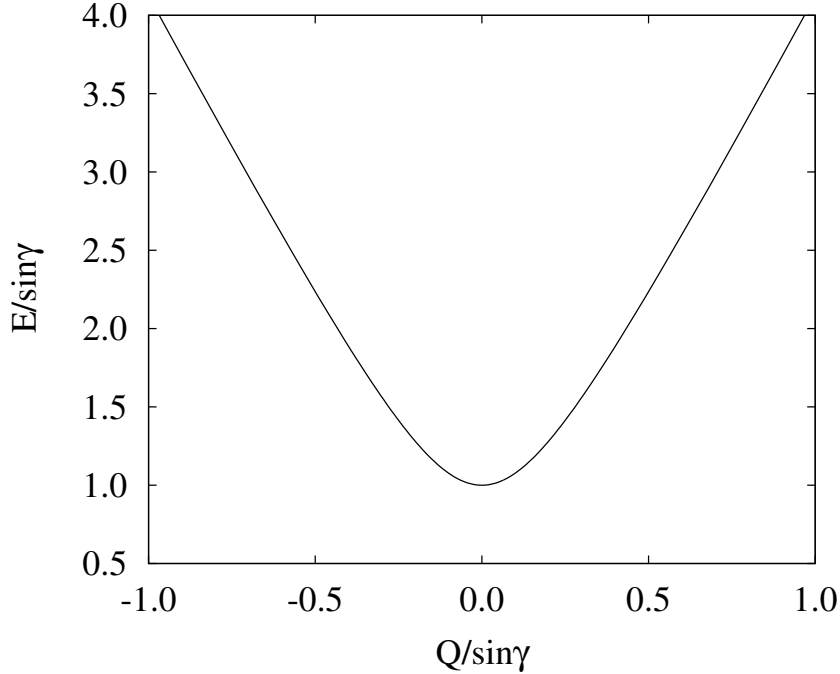}
\caption{Energy $E$ versus electric charge $Q$ for type I solutions with $\la=0$, $\ka_1=0$, and $\ka_2=0$; $\tilde V_0$ is varied and $\ga$ is kept fixed.}
\end{center}
\end{figure}

The situation changes radically when the new CS terms are present. In that case the solution with the lowest energy is not the electrically neutral one, in general. There are regions where the energy is a decreasing function of $|Q|$. Both types of HCS terms give rise to such an effect, although the first one, Eq.~\re{CS1}, requires the presence of a non-vanishing potential (i.e., $\la \neq 0$). This is shown in Fig.~2, where we exhibit the energy $E$ versus the electric charge $Q$ for type I solutions with $\tilde V_0=0$, $\ka_1=1.0$, $\ka_2=0$ and $\la=0.0$, $0.1$, and $1.0$. Clearly, the solution with the largest energy corresponds to the electrically uncharged one (excluding the vacuum solution).

\begin{figure}[ht]
\begin{center}
\includegraphics[width=0.7\textwidth]{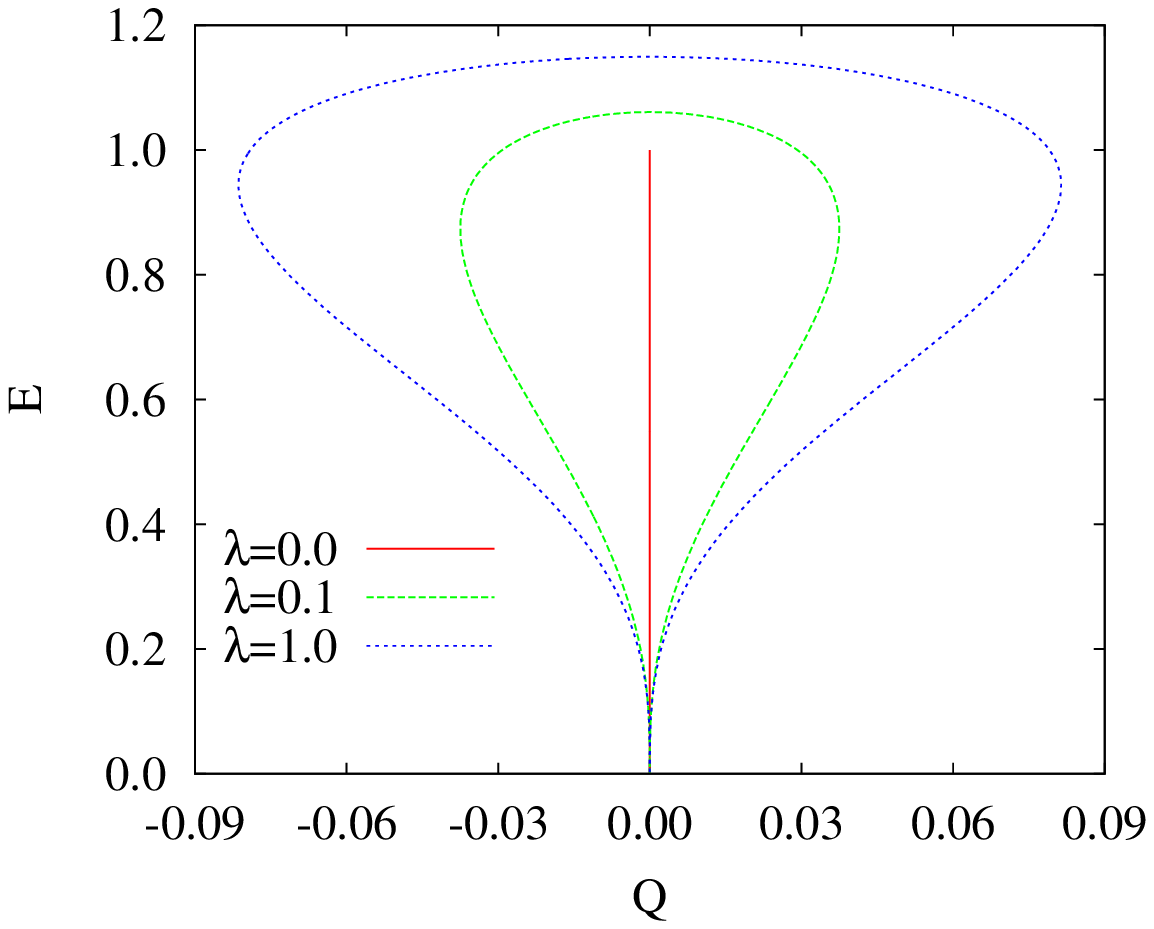}
\caption{Energy $E$ versus electric charge $Q$ for type I solutions with $\tilde V_0=0$, $\ka_1=1.0$, and $\ka_2=0$ for three values of $\la$: 0.0, 0.1, and 1.0.}
\end{center}
\end{figure}

When both contributions to the electric charge are present, the structure of the solutions gets more complicated: several solutions may exist for the same value of the electric charge. Moreover, the uncharged solutions may not exist for large enough values of $\tilde V_0$. This is exemplified in Fig.~3 where no electrically neutral solutions exist for these values of the parameters.  

\begin{figure}[ht]
\begin{center}
\includegraphics[width=0.7\textwidth]{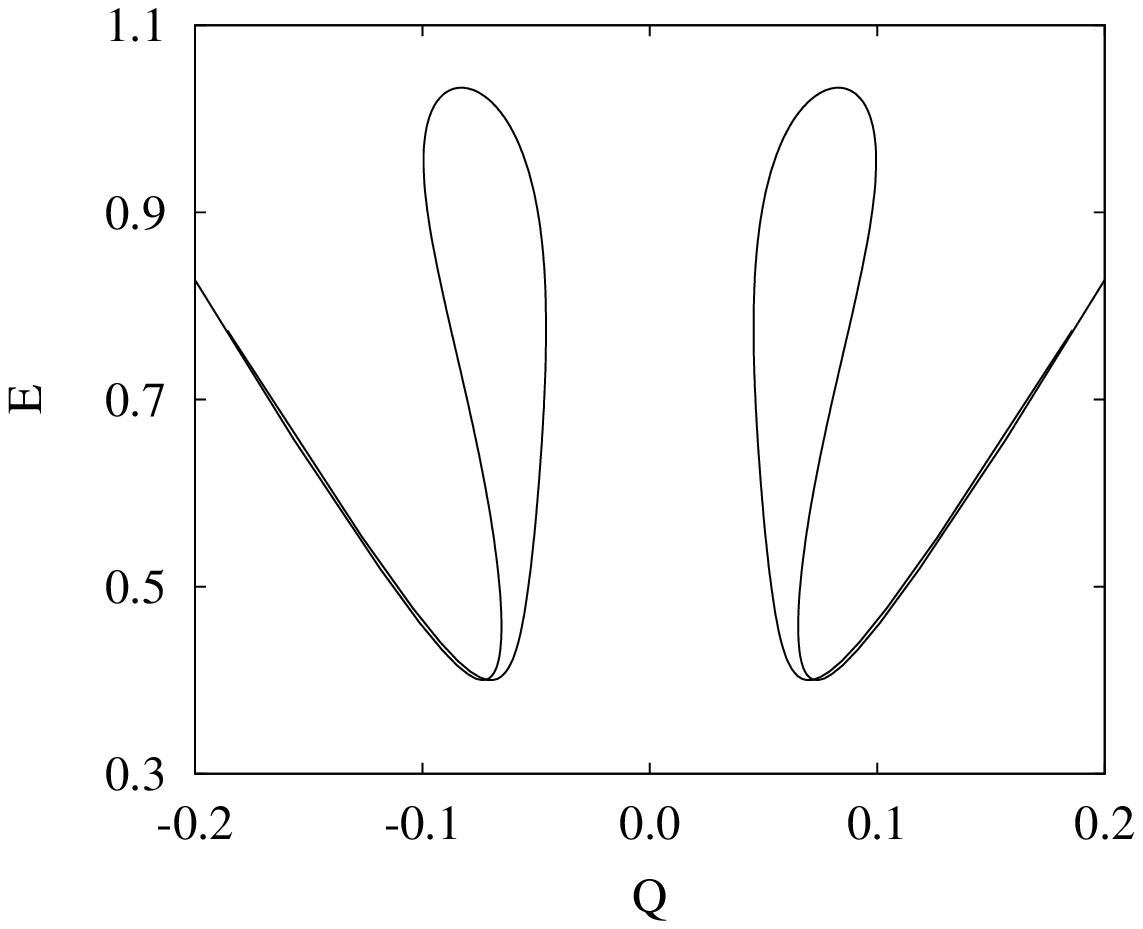}
\caption{Energy $E$ versus electric charge $Q$ for type I solutions with  $\la=0$, $\tilde V_0=0.2$, $\ka_1=0.3$, and $\ka_2=0.5$.}
\end{center}
\end{figure}

The pattern of solutions may develop a large number of branches in certain regions of the parameter space. In Fig.~4 we present the dependence of the energy $E$ on the electric charge $Q$ for type I solutions with $\tilde V_0=0.5$, $\ka_1=2.0$, $\ka_2=-12$ and $\la=0.0$. We observe that several electrically uncharged solutions exist, none of them having the lowest energy.

\begin{figure}[ht]
\begin{center}
\includegraphics[width=0.7\textwidth]{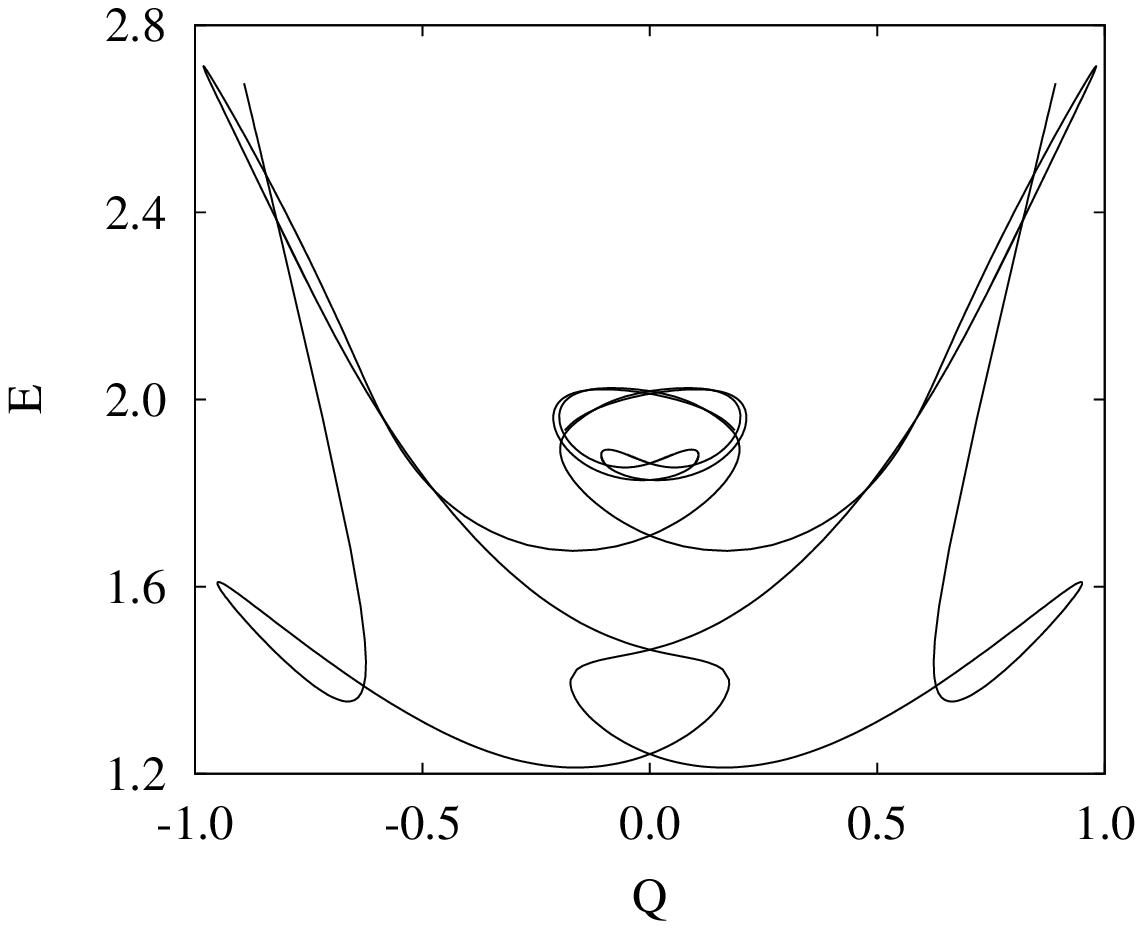}
\caption{Energy $E$ versus electric charge $Q$ for type I solutions with  $\la=0$, $\tilde V_0=0.5$, $\ka_1=2.0$, and $\ka_2=-12.0$.}
\end{center}
\end{figure}

 \subsubsection{Type II solutions}

In this case, the solutions are characterized by $\tilde w = 0$, $\tilde V=0$, $\tilde h=0$, and $\tilde g=0$. The expansions at the origin are

\bea
&&w = -1 + w_2 x^2 + 2 w_2 x^3 + O(x^4) \, , \\
&&V = {\hat V}_0 + O(x^2) \, , \\
&&g = g_1 x +g_1 x^2+ O(x^3) \, ,\\
&&h = h_0 + O(x^2) \, ,  \label{exp_or_type_II}
\eea
where $x= r/(1+r)$. The asymptotic values of the functions are
\bea
&&w = 0 \, , \\
&&V = V_0 \, , \\
&&g =  \cos\gamma \, ,\\
&&h = \sin \gamma \, , \label{exp_inf_type_II}
\eea
where $\gamma$ is free. $V$ does not enter the equations directly, but just through its derivatives. So the asymptotic value of $V$, $V_0$, may be given any arbitrary value (gauge freedom). So for this type of solutions we do not have $V_0$ as a true physical parameter to be varied; that means there is no JZ parameter. Then, only $\gamma$ allows us to vary the electric charge of the solutions, once the other parameters of the theory, namely, $\lambda$, $\kappa_1$, and $\kappa_2$, are given. 

\begin{figure}[ht]
\begin{center}
\includegraphics[width=0.7\textwidth]{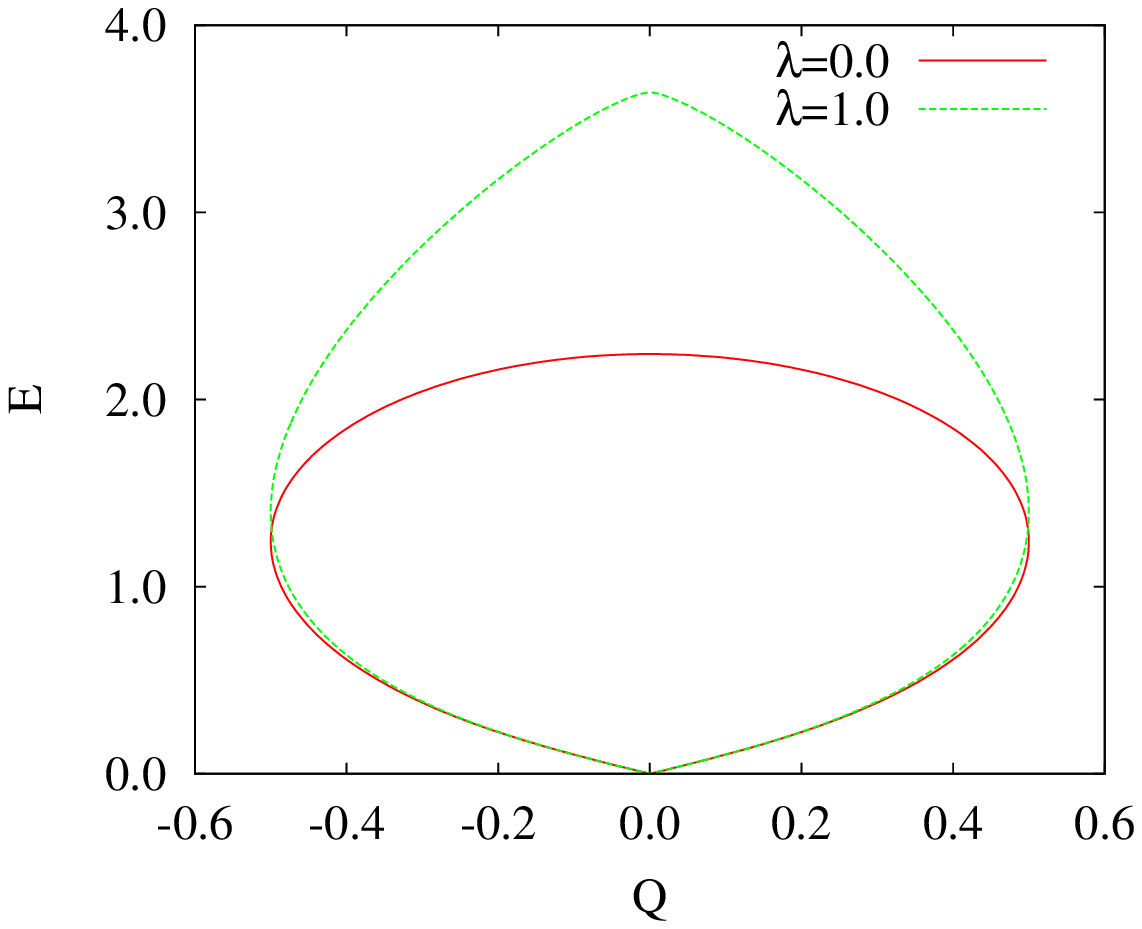}
\caption{Energy $E$ versus electric charge $Q$ for type II solutions with  $\la=0.0$, $1.0$, $\ka_1=1.0$, and $\ka_2=0.0$.}
\end{center}
\end{figure}

As opposed to type I solutions, for type II solutions the first HCS term, Eq.~\re{CS1}, can give rise to charged solutions also for $\la=0$. When only one type of the HCS term is present, the structure of the solutions is quite simple, as shown in Fig.~5. When both are present, the structure becomes more involved, although the lack of a JZ term prevents the appearance of very complicated structures as in Fig.~4. In Fig.~6 we show the energy $E$ versus the electric charge $Q$ for $\la=0.0$, $\ka_1=2.0$, and $\ka_2=-12.0$. Again, the uncharged solutions (excluding the vacuum) do not correspond to the solutions with the lowest energy.

\begin{figure}[ht]
\begin{center}
\includegraphics[width=0.7\textwidth]{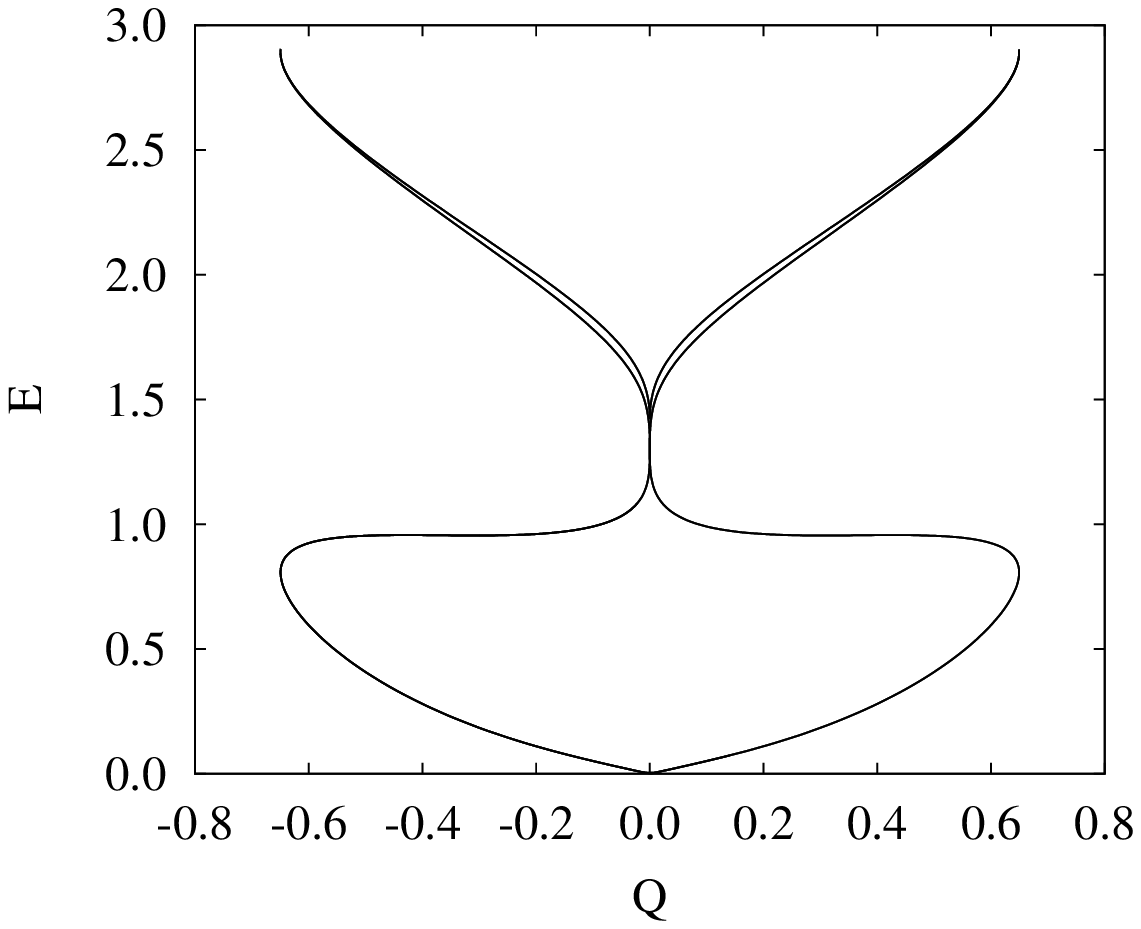}
\caption{Energy $E$ versus electric charge $Q$ for type II solutions with  $\la=0.0$, $\ka_1=2.0$, and $\ka_2=-12.0$.}
\end{center}
\end{figure}

\subsubsection{Type III solutions}

When we set $\tilde w=0$, $\tilde V=0$, $\tilde g=0$, and $h=0$, type III solutions are obtained. The expansions at the origin now read

\bea
&&w = -1 + w_2 x^2 + 2 w_2 x^3 + O(x^4) \, , \\
&&V = {\hat V}_0 + O(x^2) \, , \\
&&g = g_1 x +g_1 x^2+ O(x^3) \, ,\\
&&{\tilde h} = {\tilde h}_1 x +{\tilde h}_1 x^2+ O(x^3) \, ,  \label{exp_or_type_III}
\eea
where $x= r/(1+r)$. The asymptotic values of the functions are
\bea
&&w = 0 \, , \\
&&V = 0 \, , \\
&&g =  \cos\gamma \, ,\\
&&{\tilde h} = \sin \gamma \, , \label{exp_inf_type_III}
\eea
where $\gamma$ is free. When the electric charge $Q$, Eq.~\re{elec1}, is evaluated for these solutions, it is found to be zero. However, the electric potential, $A_0$, is not identically zero. This is clearly seen in Fig.~7 where the functions $w$, $V$, $g$, and $\tilde h$ are shown for the type III solution with $\la=0.0$, $\ga=1.2$, $\ka_1=1.0$, and $\ka_2=2.0$.

\begin{figure}[ht]
\begin{center}
\includegraphics[width=0.7\textwidth]{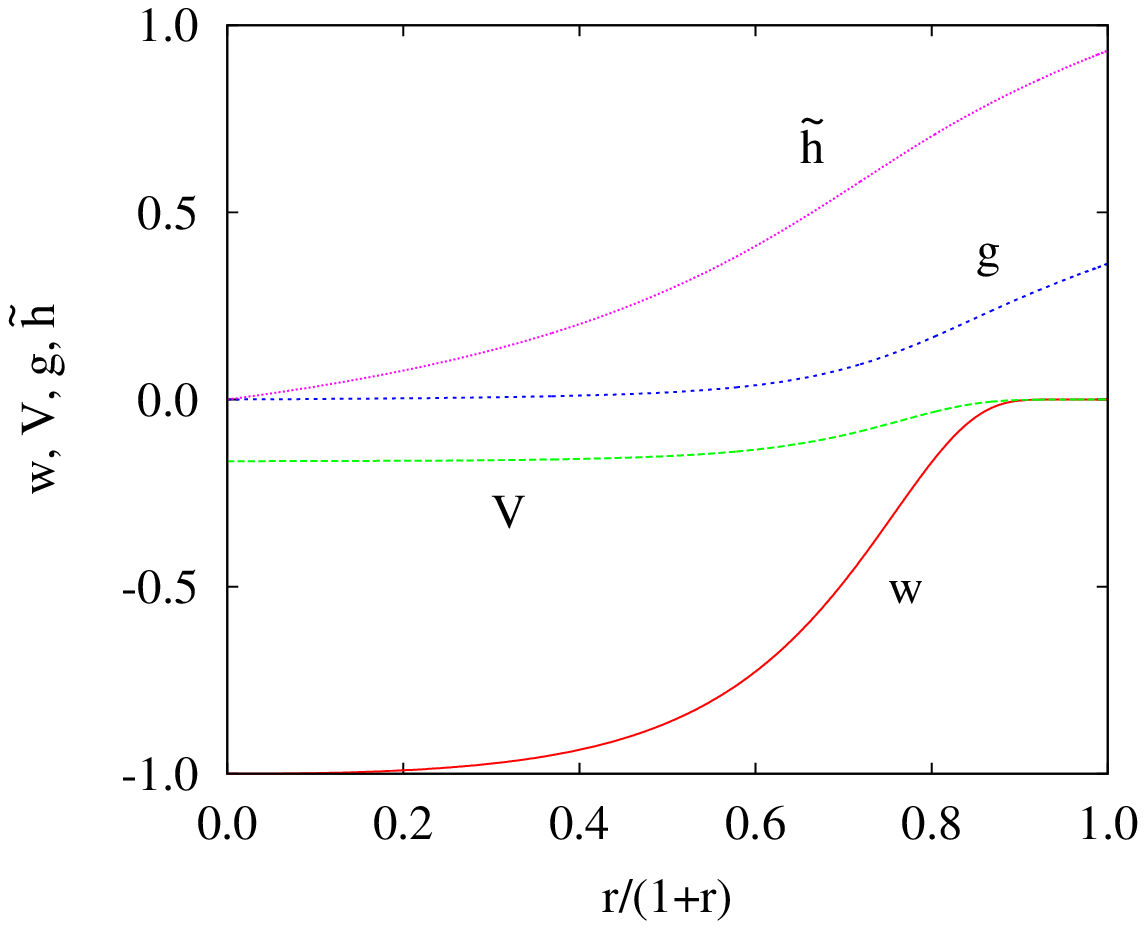}
\caption{Functions $w$, $V$, $g$, and $\tilde h$ for type III solutions with $\la=0.0$, $\ga=1.2$, $\ka_1=1.0$, and $\ka_2=2.0$. }
\end{center}
\end{figure}

Since the electric charge vanishes in this case, we may show the structure of branches plotting the energy $E$ versus the asymptotic angle $\ga$. Very intricate patterns appear, as demonstrated in Fig.~8 for the type III solutions with $\la=0.0$, $\ka_1=1.0$, and $\ka_2=-12.0$.

\begin{figure}[ht]
\begin{center}
\includegraphics[width=0.7\textwidth]{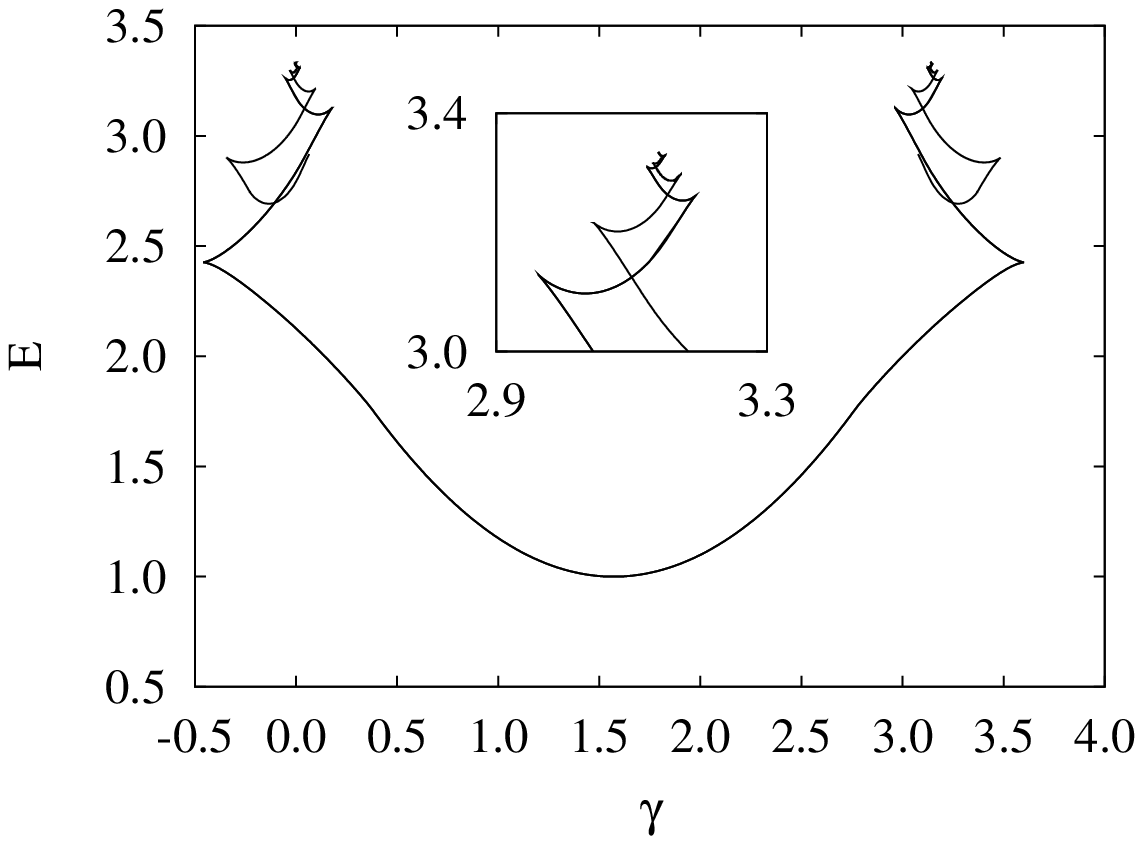}
\caption{Energy $E$ versus asymptotic angle $\gamma$ for type III solutions with  $\la=0.0$, $\ka_1=1.0$, and $\ka_2=-12.0$.}
\end{center}
\end{figure}

\section{Solutions of the $SU(3)$ Higgs-Chern-Simons--Yang-Mills-Higgs model}
This Section consists of three Subsections. In Subsection {\bf 4.1}, spherical symmetry is imposed and the boundary values of the solutions sought are
stated. The numerical construction of the solutions is presented in Subsection {\bf 4.2}, and in Subsection {\bf 4.3} we impose axial symmetry on this
system with a view to show whether the dyon of the $SU(3)$ HCS-YMH model rotates or not.

Solutions of $SU(3)$ monopoles have been studied intensively a long time ago~\cite{Corrigan:1975hd}. Here we follow the (some of the) constructions to be found in~\cite{Sinha:1976bw}
and \cite{Weinberg}.

While in the previous example, namely the $SO(5)$ model on $\R^{3+2}$, the dimensional descent from $8$ (and resly. $6$) over $S^3$
(and resly. $S^1$) giving rise to $HCS(2)$ (and resly. $HCS(1)$) was that prescribed in \cite{Tchrakian:2010ar}, here the
corresponding prescription is slightly different. Instead of the gauge field in the bulk being a $8\times 8$
(and resly. $4\times 4$) anti-Hermitian connection, here it is a $6\times 6$  (and resly. $3\times 3$) anti-Hermitian connection.

\subsection{Imposition of symmetry and boundary values}
We use the standard $SU(3)$ spherically symmetric Ansatz
\bea
A_i&=&\left(\frac{1-w}{r}\right)\,\la^{(1)}_{ij}\,\hat{x}_j \, ,\label{ai}\\
\F&=&\frac12\,i\,\eta\left(h\,\hat{x}_j\,\la^{(1)}_{j}+g\,\la_8\right)\, ,\label{h}\\
A_0&=&\frac12\,i\,\left(u\,\hat{x}_j\,\la^{(1)}_{j}+v\,\la_8\right)\, .\label{a0}
\eea
$\la^{(1)}_i$, $i=1,2,3$ are the first three $su(2)$ embeddings in $su(3)$, $\la_8$ is the last diagonal one, and
\[
\la^{(1)}_{ij}=-\frac14\,[\la^{(1)}_{i},\la^{(1)}_{j}]\,.
\]
(We have used anti-Hermitian representations of the $su(3)$ algebra.)

Detailed one dimensional reduced quantities used in our computations are given in Appendix {\bf A.2}.

The expansions at the origin read
\bea
&&w = -1 + w_2 x^2 + 2 w_2 x^3 + O(x^4) \, , \\
&&h = h_1 x + h_1 x^2 + O(x^3) \, , \\
&&g = g_0+ O(x^2) \, ,\\
&&u = u_1 x + u_1 x^2  + O(x^3) \, , \\
&&v = v_0 + O(x^2) \, .  \label{exp_or_su}
\eea

We seek solutions with the following asymptotic values

\bea
\lim_{r\to 0}\,w(r)&=&1\ ,\quad\lim_{r\to \infty}\,w(r)=0 \, , \label{asymw_su}\\
\lim_{r\to 0}\,h(r)&=&0\ ,\quad\lim_{r\to \infty}\,h(r)=\cos\ga \, , \label{asymh_su}\\
\lim_{r\to 0}\,g'(r)&=&0\ ,\quad\lim_{r\to \infty}\,g(r)=\sin\ga \, ,\label{asymg_su}\\
\lim_{r\to 0}\,u(r)&=&0\ ,\quad\lim_{r\to \infty}\,u(r)=u_0 \, , \label{asymu_su}\\
\lim_{r\to 0}\,v'(r)&=&0 \ ,\quad\lim_{r\to \infty}\,v(r)=0 \text{  (gauge choice)}\, ,  \label{asymv_su}
\eea
where $\ga$ and $u_0$ are free parameters, corresponding to an asymptotic angle for the Higgs components and the JZ parameter, respectively.

Under these boundary conditions, the magnetic charge, Eq.~\re{tH-P}, becomes
\be
\mu = \eta \cos\ga \label{mu_val} \, ,
\ee
and the electric charge, Eq.~\re{elec}, results to be
\be
\label{elec1_su2}
Q=\frac12 \eta \,\left[r^2\,(hu'+gv')\right]_{r=\infty} \, .
\ee

In the absence of the HCS terms, when $\la=0$ the second-order field equations are solved by the first-order
selfduality equations. The latter
reduce to the BPS equations which have nontrivial solutions only for the functions $w(r)$, $h(r)$ and $u(r)$,
while the functions $g(r)$ and $v(r)$ both vanish everywhere. This means that with $\la=0$ the only solutions are the $SU(2)$ JZ
dyons in that case.
However, when the HCS terms are present, nontrivial solutions for the functions $g(r)$ and $v(r)$ are present even in the $\la=0$ limit. Since the parameter space is already large enough, we will restrict our attention in this work to the $\la=0$ case only, for economy of presentation.

\subsection{Numerical results}

We have generated numerical solutions to this theory. In these numerical ruesults we have set $\eta=1$ to fix the scale.  



As happened for $SO(5)$, when $\kappa_1=0$, $\kappa_2=0$, and $\lambda=0$ the representation of the scaled energy $E/\mu$ versus the scaled electric charge $Q/\mu$ shows that $E/\mu$ is an increasing function of $|Q/\mu|$; in fact, the figure coincides with Fig.~1 (when rescaled properly). 
The situation changes, however, when the HCS terms are present. In that case, for a given asymptotic angle $\ga$ (i.e., a given magnetic charge $\mu$), the electrically uncharged solution need not be the one with the least energy. We exhibit this fact in Fig.~9 where we represent the energy $E$ versus the magnetic charge $\mu$ for $\la=0$, $\ka_1=1$, and $\ka_2=1$ and several values of the electric charge Q: 0.0, 0.5, and 1.0. (Notice that in the limit $\mu=0$ the value of the energy tends to the value of the electric charge.)

\begin{figure}[h]
\begin{center}
\includegraphics[width=0.7\textwidth]{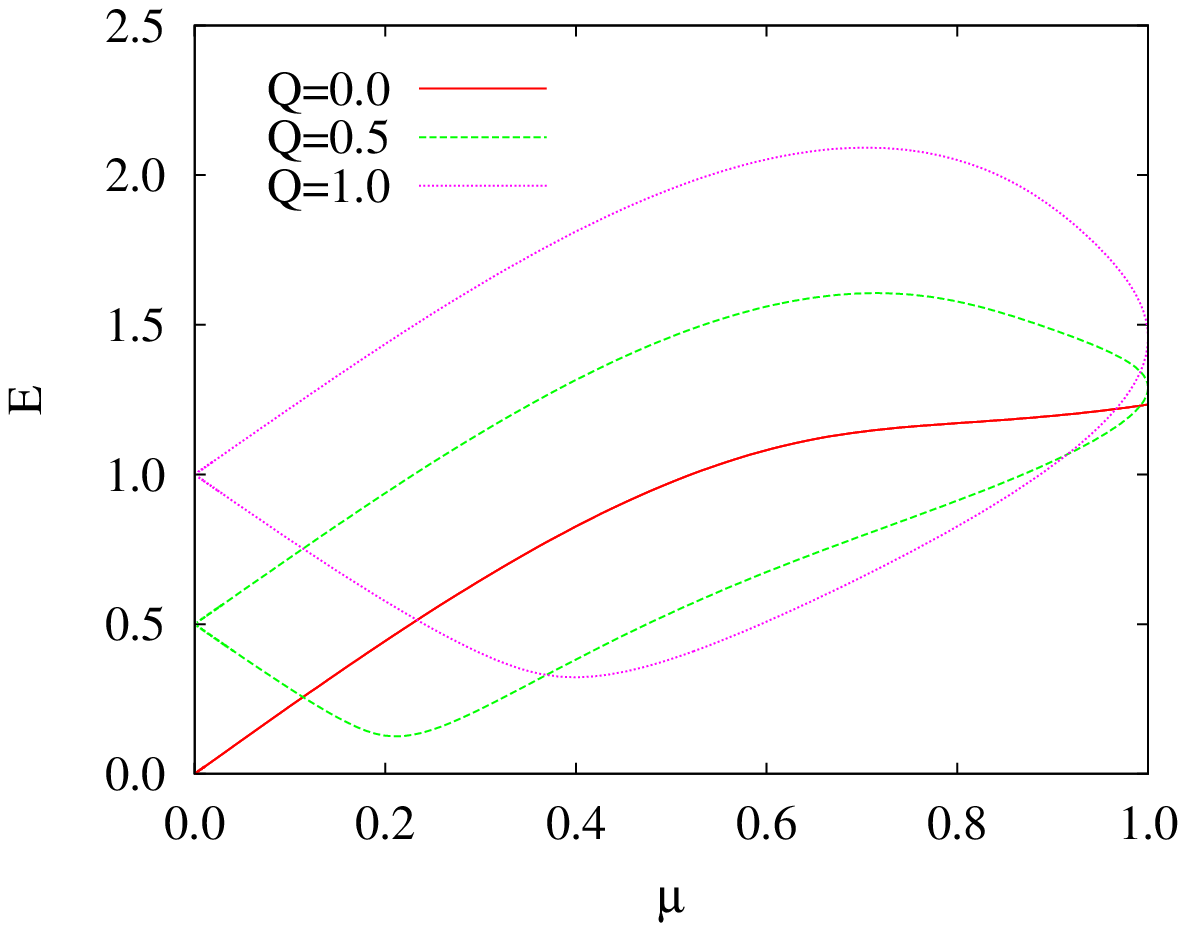}
\caption{Energy $E$ versus magnetic charge $\mu$ for solutions with $\la=0$, $\ka_1=1$, and $\ka_2=1$ and several values of the electric charge $Q$: 0.0, 0.5, and 1.0.}
\end{center}
\end{figure}

\begin{figure}[h]
\begin{center}
\includegraphics[width=0.7\textwidth]{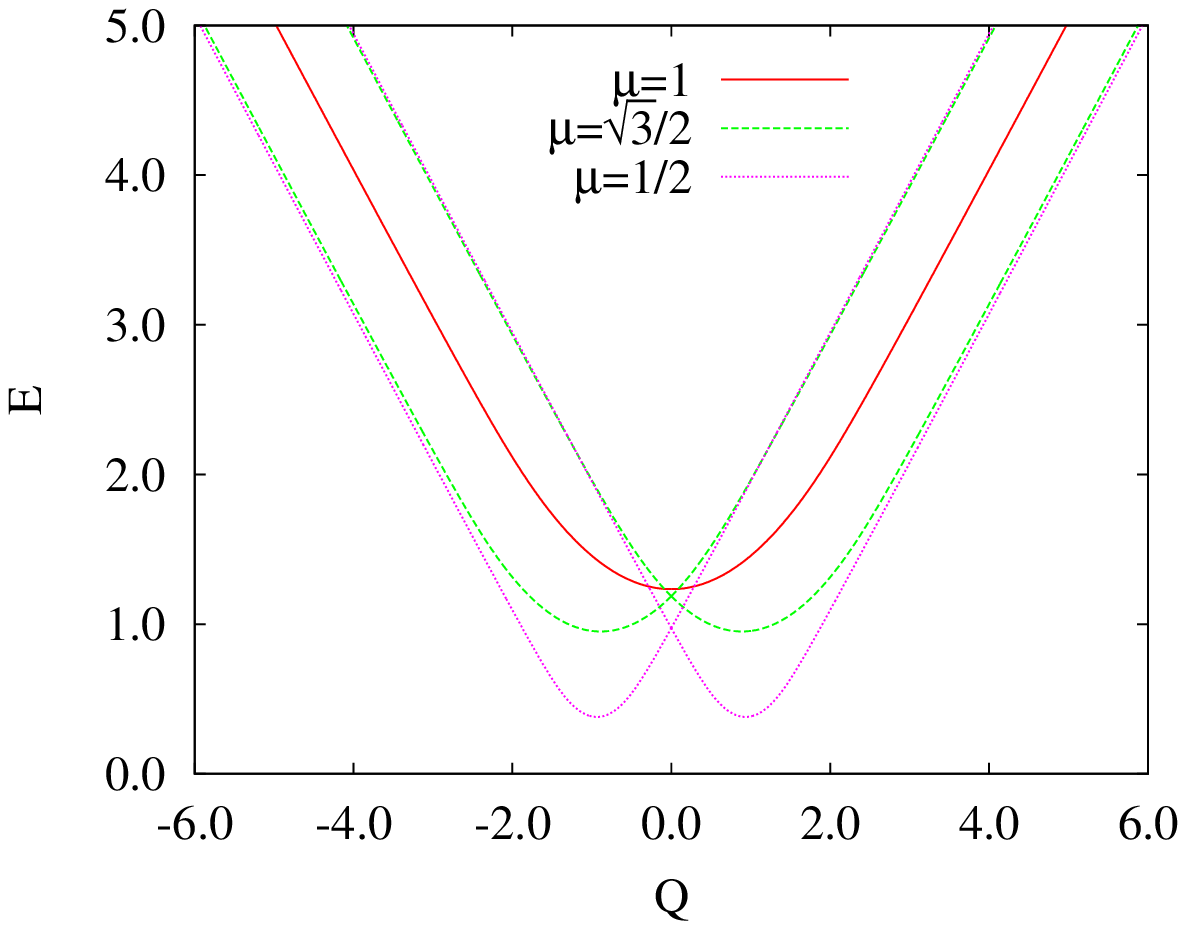}
\caption{Energy $E$ versus electry charge $Q$ for solutions with $\la=0$, $\ka_1=1$, and $\ka_2=1$ and several values of the magnetic charge $\mu$: $1$,$ \sqrt{3}/2$, and $1/2$.}
\end{center}
\end{figure}

This effect is more clearly observed in Fig.~10, where we represent the energy $E$ of the solutions versus the electric charge $Q$ for 3 asymptotic angles $\ga=0$, $\pi/6$, and $\pi/3$ for $\la=0$, $\ka_1=1$, and $\ka_2=1$. For nonvanishing $\ga$ the minimal energy occurs for a nonvanishing of the electric charge.

\subsection{The issue of angular momentum}
The issue of angular momentum density can readily be calculated using the Ansatz given in Eqs.~\re{Aal}-\re{F},
\be
\label{Tf01}
4\,T_{\vf}{}^0=\left[(D_{\rho}\xi\cdot D_{\rho}\chi)+(D_{z}\xi\cdot D_{z}\chi)\right]+4\eta^2\rho\,(\f\vep\chi)(\f\vep\xi) \, ,
\ee
which can be rewritten in the form
\bea
\label{Tf02}
4\,\rho\,T_{\vf}{}^0&=&\left[\pa_{\rho}(\rho\,\xi\cdot D_{\rho}\chi)+\pa_{z}(\rho\,\xi\cdot D_{z}\chi)\right]\nonumber\\
&&-\left[(\xi\cdot D_{\rho}\chi)+\rho\,\xi\cdot(D_{\rho}D_{\rho}\chi+D_{z}D_{z}\chi)\right]-4\eta^2\rho\,(\f\vep\chi)(\f\vep\xi) \ ,
\eea
where a total divergence term is isolated.

Consider now the equation resulting from the variation of Eq.~\re{redHCS1} with respect to the doublet $\chi^a$,
\bea
\label{elchi}
&&\left[D_{\rho}\chi^a+\rho(\,D_{\rho}D_{\rho}\chi^a+D_{z}D_{z}\chi^a)-\frac1\rho(\chi\vep\xi)(\vep\xi)^a\right]+4\eta^2\rho\,(\f\vep\chi)(\f\vep)^a=\nonumber\\
&&\qquad\qquad\qquad\qquad\qquad\qquad=
\frac12\,\ka_1\,\eta\left[D_{[\rho}(g\,D_{z]}\xi)^a+g\,f_{\rho z}\,(\vep\xi)^a\right] \, .
\eea
Contracting Eq.~\re{elchi} with $\xi^a$ and substituting the result in Eq.~\re{Tf02}
\be
\label{J}
4\,\rho\,T_{\vf}{}^0=\,\left[\pa_{\rho}(\rho\,\xi\cdot D_{\rho}\chi)
+\pa_{z}(\rho\,\xi\cdot D_{z}\chi)\right]-\frac12\,\ka_1\,\eta\,\left[\xi\cdot D_{[\rho}(g\,D_{z]}\xi)\right] \, .
\ee
The first term on right-hand side in Eq.~\re{J} is a {\bf div} and its volume integral vanishes by virtue of the asymptotic values of the solutions.


The second term is a {\bf curl}. Using the notation
\[
x_A=(\rho,\,z)\,,
\]
the second term in Eq.~\re{J} can be expressed as
\bea
\label{curl}
\xi^a\, D_{[\rho}(g\,D_{z]}\xi)^a&=&\vep_{AB}\,\xi^a\, D_{A}(g\,D_{B}\xi)^a\nonumber\\
&=&-\frac12\,\vep_{AB}\,\pa_A\left(|\xi|^2\,\pa_Bg\right)\nonumber \, ,
\eea
which can be evaluated by performing a contour integral, using Stokes' Theorem (like the multi-monopole charge.)

On the far hemisphere, $|\xi|^2=0$ so there will be no contribution. On the $z$-axis $\pa_zg$ changes sign going through the origin,
so the line integral on the positive $z$-axis will cancel against the line integral on the negative $z$-axis. Thus, the angular
momentum of this system vanishes.

\section{Summary, comments and outlook}
In this Section, we will summarise our results and comment on their properties. After that we will describe what further questions may arise out of the results.
In this paper we have constructed electrically charged solitons in two distinct YMH models in $3+1$ dimensions, one with gauge group $SO(5)$ and the other $SU(3)$. Both these
theories involve two (dynamical) new CS terms which we refer to as HCS terms. The purpose of this investigation is to show that in certain regions of the parameter
space, the electrically charged solutions have smaller mass than their electrically neutral counterparts. This property is a consequence of the dynamics of the
HCS densities appearing in the respective Langrangian. This is the main result presented here.

This investigation is carried out for two distinct models to show that the main result obtained here, is independent of the specific feature of the model chosen, namely of the choice
of gauge group. The $SO(5)$ and $SU(3)$ models employed differ in an important respect, namely that the former has zero magnetic charge while the latter has a magnetic charge
(in the spherically symmetric case). It is reasonable to treat these two types of solutions separately, to ensure that such a prominent difference does not result in the main feature claimed.

Solutions to the $SO(5)$ and $SU(3)$ models share two properties. First, when the HCS terms are decoupled, $i.e.$ setting $\ka_1=\ka_2=0$, the energy of the charged soliton
increases with increasing electric charge. This expected result is exhibited in Figure. $1$.

Another consequence of setting $\ka_1=\ka_2=0$ in these models is, that in the absence of the Higgs symmetry breaking potential ($\la=0$) only solutions parametrising the $SO(3)$ subgroup are supported.
However, when $\ka_1$ and/or $\ka_2$ are switched on, the gauge fields can take their values outside of $SO(3)$. It is therefore not necessary to consider $\la>0$ solutions and for simplicity we have
concentrated on the $\la=0$. We have nonetheless considered $\la>0$ models in a few cases, to ensure that the introduction of the Higgs potential does not alter the qualitative features of our
main result. 

\subsection{The $SO(5)$ model}
In this case we have only
zero magnetic charge solutions. These exhibit the desired property in some regions of the parameter space. To make our investigation complete, we have studied three
types of such solutions, Type I, II and III. The numerical construction of these solutions is presented in Section {\bf 3.2}.
\begin{itemize}
\item
Type I solutions are characterised by the existence of two parameters: one of them related to the JZ contribution to the electric charge, $\tilde V_0$, and one related to the HCS contribution, $\gamma$. These solutions posses a non-vanishing electric charge coming from both types of sources. Uncharged solutions may have higher energy than the charged ones. These results are exhibited in Figs.~1-4 where we exhibit the dependence of the energy $E$ on the electric charge $Q$ under several circumstances.
\item
Type II solutions are characterised by the presence of the asymptotic angle $\gamma$. In this case there is no JZ parameter free.  These results are exhibited in Fig.~5 and 6. As for Type I solutions, these solutions are electrically charged and their mass may be lower than that of the uncharged solution. 
\item
Type III solutions are characterised also by the asymptotic angle $\gamma$. Opposite to the previous two type these solutions are electrically uncharged although their electric potential is not identically zero. They describe electric dipoles with zero electric monopole. These results are exhibited in Figs.~7 and 8. The structure of these solutions may get quite complicated as shown in Fig.~8. 
\end{itemize}
Note that in Figs~2 and 5, profiles with $\la>0$ appear, which preserve the shapes conformally. 

\subsection{The $SU(3)$ model}
The main feature in this case is that the solutions carry both electric and magnetic charge, and are dyons. We see that the qualitative features observed in the $SO(5)$ model, namely our main
result, are preseved. While the qualitative result, that the electrically neutral solutions can be more massive than the neutral ones, a specific feature is observed.
\begin{itemize}
\item
In Fig.~9 we observe that for non-vanishing electric charge, two dyonic solutions are possible for a magnetic charge $0<\mu<1$. The mass of the magnetic monopole (curve in red) is higher than the corresponding value along the lower branch for large ranges in $\mu$. That indicates that the electrically neutral solutions are not necessarily the least energetic ones, in general. 
\item
In Fig.~10 we show this effect more clearly for 3 values of the magnetic charge (including the one chosen in Weinberg's book \cite{Weinberg} (green curve)). For $0<\mu<1$ the minimum of the energy occurs for non-vanishing electric charge.
\end{itemize}
In addition in this case we have considered the axially symmetric fields and have constructed the angular momentum density of this $SU(3)$ dyon. It turns out that this vanishes.

\subsection{Summary and outlook}
In this paper we have constructed electrically charged solitons in an $SO(5)$ and $SU(3)$ HCS-YMH theory in $3+1$ dimensions. These theories contain the new CS terms which were employed in \cite{Navarro-Lerida:2013pua} for the $SO(5)$ model. By means of an enlarged spherically symmetric Ansatz, we have been able to endow the solutions of the $SO(5)$ model~\cite{Navarro-Lerida:2013pua} with an asymptotic angle $\ga$ resulting in a larger set of electrically charged solutions, which exhibit the new desired properties. Qualitatively similar results are obtained for the $SU(3)$ model. This way of producing electrically charged solutions differs from the prescription of Julia and Zee~\cite{Julia:1975ff}. Technically, in the $SO(5)$ model, the obvious difference with the JZ prescription is that the time component of the YM potential $A_0$ and the Higgs field do not take their values in the same representation of the gauge group. But more importantly, the origin of the electrical fields here is found in the CS dynamics in the case of both the $SO(5)$ and $SU(3)$ models. This is akin to the analogous $2+1$ dimensional situation in \cite{Paul:1986ix} and \cite{Hong:1990yh,Jackiw:1990aw}.



In the case of the $SU(3)$ model we have calculated the angular momentum of the CS dyon and found that it vanishes. In this respect, the introduction of a new CS term with the attendant enlargement of the gauge group from $SU(2)$ to $SU(3)$, does not change the general result in \cite{Navarro-Lerida:2014bja} (and references therein), namely that $SU(2)$ YMH dyons in $3+1$ dimensions do not rotate. This property contrasts with the analogous $2+1$ dimensional situation in \cite{Paul:1986ix} and \cite{Hong:1990yh,Jackiw:1990aw}, where the introduction of the CS term results in rotation. In the matter of electric charge the introduction of a CS term plays the same role in gauge-Higgs theories in both $3+1$ and $2+1$ dimensions. Thus, the effect of CS dynamics in $3+1$ and $2+1$ dimensions is qualitatively different, overlapping in one respect (electric charge) but differing in another (angular momentum). This question is at present under intensive consideration.

Finally, it is natural to inquire what the analogue of the present investigation in the context of gauged Higgs models would be, in the case of gauged Skyrme~\cite{Skyrme:1962vh} systems. For this, one would have to employ the Skyrme analogue of the HCS densities used here. This question is at also under intensive consideration.

\bigskip

{\bf Acknowledgments}
We thank Eugen Radu for fruitful discussions and suggestions on this paper. 
D.H.Tch. thanks Hermann Nicolai for his hospitality at the Albert-Einstein-Institute, Golm,
(Max-Planck-Institut, Potsdam) where parts of this work were carried out. 
F. N-L. acknowledges financial support of the Spanish Education and Science Ministry under Project No. FIS2011-28013 (MINECO).

\appendix
\section{The one dimensional quantities subject to spherical symmetry}
\setcounter{equation}{0}
\renewcommand{\theequation}{A.\arabic{equation}}
In this Appendix, we present the curvature field strengths and the covariant derivatives subject to spherical symmetry. The resulting
one dimensional static Lagrangian and energy densities used in our computations are then displayed. These quantities are given in the
following two subsections, each for the $SO(5)$ and the $SU(3)$ models, respectively.

\subsection{$SO(5)$ model}
The parametrisation used in the Ansatz, Eqs.~\re{higgs}-\re{aip}, results in a gauge
covariant expression for the YM curvature $F_{\mu\nu}=(F_{ij},F_{i0})$ and the covariant derivative of the Higgs $D_{\mu}\F=(D_{i}\F,D_{0}\F)$
\bea
F_{ij}&=&\frac{1}{r^2}\left(|\vec\xi|^2-1\right)\Sigma_{ij}+
\frac1r\left[D_r\xi^{6}+\frac1r\left(|\vec\xi|^2-1\right)\right]
\hat x_{[i}\Sigma_{j]k}\hat x_{k}+
\frac1rD_r\xi^M\hat x_{[i}\Sigma_{j]M}\, ,\label{fij}\\
F_{i0}&=&-\frac1r\,\xi^M(\vep\chi)^M\,\Sigma_{ij}\hat x_j+\frac1r\,
\left[\xi^{6}(\vep\chi)^M-\chi^{6}(\vep\xi)^M\right]\Sigma_{iM} 
\nonumber\\
&&-\left\{(\vep D_r\chi)^M+\frac1r\,
\left[\xi^{6}(\vep\chi)^M-\chi^{6}(\vep\xi)^M\right]\right\}
\hat x_i\hat x_j\Sigma_{jM}
-D_r\chi^{6}\,\hat x_i\,\Sigma_{45}\, , \label{fi0}\\
(2\eta)^{-1}D_i\F&=&-\frac1r(\vec\xi\cdot\vec\f)(\delta_{ij}-\hat{x_i}\hat{x_j})\,\Si_{j6}
+D_r\f^M\hat{x_i}\,\Si_{M6}+D_r\f^6\,\hat{x_i}\hat{x_j}\,\Si_{j6}\nonumber\\
&&-\frac1r\,\xi^M(\vep\psi)^M\,\Sigma_{ij}\hat x_j+\frac1r\,
\left[\xi^{6}(\vep\psi)^M-\psi^{6}(\vep\xi)^M\right]\Sigma_{iM}
\nonumber\\
&&-\left\{(\vep D_r\psi)^M+\frac1r\,
\left[\xi^{6}(\vep\psi)^M-\psi^{6}(\vep\xi)^M\right]\right\}
\hat x_i\hat x_j\Sigma_{jM}
-D_r\psi^{6}\,\hat x_i\,\Sigma_{45}\, , \label{DiF}\\
(2\eta)^{-1}D_0\F&=&\f^M(\vep\chi)^M\hat x_j\,\Sigma_{j6}-\left[\f^{6}(\vep\chi)^M
-\chi^{6}(\vep\f)^M\right]\Sigma_{M6}\nonumber\\
&&+\chi^{M}\psi^{N}\,\Si_{MN}-(\psi^6\chi^M-\chi^6\psi^M)\,\hat{x}_j\,\Si_{jM} \, ,
\label{D0F}
\eea
in which we have used the notation
\bea
D_r\f^a&=&\pa_r\f^a+\vep^{abc}\,A_r^b\,\f^c\ , \quad ...\nonumber
\eea
as the $SO(3)$ covariant derivatives of the four triplets
$\vec\xi\equiv\xi^a=(\xi^M,\xi^{6})$,
$\vec\chi\equiv\chi^a=(\chi^M,\chi^{6})$,
$\vec\psi\equiv\psi^a=(\psi^M,\psi^{6})$, and
$\vec\f\equiv\f^a=(\f^M,\f^{6})$, with respect to the
$SO(3)$ gauge connection $\vec A_r\equiv A_r^a$.

Substituting Eq.~\re{higgs} and Eqs.~\re{fi0} in the HCS densities, Eqs.~\re{CS1}-\re{CS2},
we have the reduced one dimensional HCS densities
\be
\label{redCSs}
\omega^{(i)}_{\rm CS} \stackrel{\rm def.}= \kappa_i r^2 \Omega^{(i)}_{\rm CS}\ ,\quad i=1,2\, ,
\ee
where for the first HCS term, Eq.~\re{CS1}, we have the reduced one dimensional density $\omega_{\rm{CS}}^{(1)}$,
\be
\label{redCS1}
\omega_{\rm{CS}}^{(1)} = 8\ka_1\,\eta\,\left[(|\vec\xi|^2-1)\,\vec\f\cdot D_r\vec\chi-
2(\vec\xi\times\vec\chi)\cdot(\vec\f\times D_r\vec\xi)\right]\,,
\ee
which does not receive a contribution from the triplet $\vec\psi$.
The second HCS term, $\omega_{\rm{CS}}^{(2)}$, Eq.~\re{CS2}, however does receive a contribution from $\vec\psi$.
The resulting expression being too cumbersome and not instructive, we do not exhibit it here. We have of course verified that
its computation using symbolic manipulations is correct.

The reduced one dimensional YM Lagrangian is
\be
\label{redYM}
-L_{\rm{YM}}^{(1)}=\left(2\,|D_r\vec\xi|^2+
\frac{1}{r^2}\left(|\vec\xi|^2-1\right)^2\right)
-\left(r^2\,|D_r\vec\chi|^2
+2\,|(\vec\xi\times\vec\chi)|^2\right)\, ,
\ee
the reduced one dimensional Higgs Lagrangian is
\bea
\label{redhiggs}
L_{\rm{Higgs}}&=&2\,\eta^2\,r^2\,\bigg\{|(\vec\f\times\vec\chi)|^2-
\left[|D_r\vec\f|^2+\frac{2}{r^2}(\vec\xi\cdot\vec\f)^2\right]\nonumber\\
&&\qquad\quad+|(\vec\psi\times\vec\chi)|^2-
\left[|D_r\vec\psi|^2+\frac{2}{r^2}(\vec\xi\times\vec\psi)^2\right]
\bigg\}\,,
\eea
and, finally, the Higgs potentials, Eqs.~\re{V_1} and \re{V_2}, reduce (for $a_1=1/4$ and $a_2=1$) to
\bea
v_1&=&\eta^4\,r^2\left[1-\left(|\vec\f|^2+|\vec\psi|^2\right)\right]^2\label{v_1}\, , \\
v_2&=&\eta^4\,r^2\left(\left[1-\left(|\vec\f|^2+|\vec\psi|^2\right)\right]^2+4(\vec\f\cdot\vec\psi)^2\right)\, , \label{v_2}
\eea
with
\[
v_i \stackrel{\rm def.}= r^2 V_i \, ,\quad i=1,2\,.
\]
It is clear that in the case Eq.~\re{v_2} the asymptotic triplet $\vec\f$ must be orthogonal to the asymptotic triplet $\vec\psi$.

Another quantity we will employ to analyze the solutions is their energy, $E$, given by
\bea
\label{energy} 
E &=& \int_0^\infty \left\{|D_r\vec\xi|^2+
\frac{1}{2r^2}\left(|\vec\xi|^2-1\right)^2
+\frac{1}{2}r^2\,|D_r\vec\chi|^2
+|(\vec\xi\times\vec\chi)|^2 \right. \nonumber\\
&&+2\,\eta^2\,r^2\,\left[|(\vec\f\times\vec\chi)|^2+
|D_r\vec\f|^2+\frac{2}{r^2}(\vec\xi\cdot\vec\f)^2
+|(\vec\psi\times\vec\chi)|^2+
|D_r\vec\psi|^2+\frac{2}{r^2}(\vec\xi\times\vec\psi)^2
\right] \nonumber \\
&&\left.+\frac{\la}{2} \eta^4\,r^2\left[1-\left(|\vec\f|^2+|\vec\psi|^2\right)\right]^2
 \right\} \, dr \, .
\eea
Notice that only the first potential Eq.~\re{V_1} has been included. 

\subsection{$SU(3)$ model}
Subject to the Ansatz Eq.~\re{h}, the symmetry breaking potentials, Eqs.~\re{V_1} and \re{V_2}, reduce, respectively, to
\bea
V_1&=&\eta^4\left[1-\left(h^2+g^2\right)\right]^2 \, ,\label{v_1_su}\\
V_2&=&\frac{1}{4}\eta^4\left[3-a_2\,(h^2+g^2)+\frac{a_2^2}{8}(h^2+g^2)^2\right]\, ,\label{v_2_su}
\eea
with $a_1=2$. It is clear that in the case Eq.~\re{v_2_su}, $V_2$ cannot vanish for any real value of the constant $a_2$, $i.e.$ we have only one choice in this case, namely Eq.~\re{v_1_su}.

The resulting curvatures and covariant derivative following from Eqs.~\re{ai}, \re{h} and \re{a0}, are
\bea
F_{ij}&=&-\frac{1}{r^2}(1-w^2)\,\la^{(1)}_{ij}-\left[\frac{w'}{r}+\frac{1}{r^2}(1-w^2)\right]\hat{x}_{[i}\,\la^{(1)}_{j]k}\,\hat{x}_k \, ,\label{fij_su}\\
D_i\F&=&\frac12\,i\,\eta\left[\frac{wh}{r}\,\la^{(1)}_{i}+\left(h'-\frac{wh}{r}\right)\hat{x}_{i}\hat{x}_{j}\la^{(1)}_{j}+g'\,\hat{x}_{i}\,\la_8\right] \, ,\label{Dif}\\
F_{i0}&=&\frac12\,i\,\left[\frac{wu}{r}\,\la^{(1)}_{i}+\left(u'-\frac{wu}{r}\right)\hat{x}_{i}\hat{x}_{j}\la^{(1)}_{j}+v'\,\hat{x}_{i}\,\la_8\right] \, ,\label{fi0_su}\\
D_0\F&=& 0  \, ,\label{D0f_su}
\eea
further resulting in
\bea
\mbox{Tr}\,F_{ij}^2&=&-\frac{1}{r^2}\left[2\,w'^2+\frac{1}{r^2}(1-w^2)^2\right]\, ,\label{fij2_su}\\
\mbox{Tr}\,F_{i0}^2&=&-\frac12\,\left[u'^2+\frac{2}{r^2}w^2u^2+v'^2\right]\, ,\label{fi02}\\
\mbox{Tr}\,D_{i}\F^2&=&-\frac12\,\eta^2\left[h'^2+\frac{2}{r^2}w^2h^2+g'^2\right]\, .\label{Dif2_su}
\eea
The magnetic charge integral, Eq.~\re{tH-P}, reduces to
\be
\label{tH-P1}
\mu=\eta[(1-w^2)h]_{r=\infty}
\, ,
\ee
and the electric charge integral, Eq.~\re{elec}, results to be
\be
\label{elec1_su}
Q=\frac12 \eta \,\left[r^2\,(hu'+gv')\right]_{r=\infty} \, .
\ee

The energy of the solutions is given by
\be
E = \frac{1}{4}\int_0^\infty \left[ r^2 u'^2 + r^2v'^2 + 2u^2w^2 + \frac{(1-w^2)^2}{r^2} + 2 w'^2 
+ r^2 g'^2 + r^2 h'^2 + 2 h^2 w^2 + 2\la r^2 (1- g^2-h^2)^2  \right] \ dr \, .
\label{energy_su}
\ee

Subject to this spherical symmetry, the HCS densities Eqs.~\re{CS1} and \re{CS2}, do not identically vanish but yield
\be
\label{CS1aa}
\Omega_{\rm CS}^{(1)}=-\frac{2}{\sqrt{3}r^2}\eta\,\left[(1-w^2)\,h\,v'+g[(1-w^2)u]'\right]\, ,
\ee
\bea
\Omega_{\rm CS}^{(2)}&=&-\frac{\sqrt{3}}{54 r^2}\eta^3 \left\{ -2 guw(36-g^2-5h^2)w'
+3 h\left[ (1-w^2)(12-g^2- h^2) + 2 h^2w^2\right]  v' \right. \nonumber \\
&& \left. + g\left[ (1-w^2)(36-g^2-9 h^2) + 2 h^2w^2\right]u'
-h^2 u w^2 g' + 4 ghuw^2 h' \right\}  . \label{CS2aa}
\eea

\section{Imposition of axial symmetry on the $SU(3)$ model}
\setcounter{equation}{0}
\renewcommand{\theequation}{B.\arabic{equation}}
In this Appendix, we present the axially symmetric field configurations employed in Section {\bf 4.3}, in the discussion of the issue of angular momentum in the $SU(3)$ model.

We denote the magnetic component $A_i=(A_{\al},A_z)$ of the $SU(3)$ connection corresponding to the spherically symmetric Ansatz Eq.~\re{ai} as
\be
\label{axai}
A_{i}=\left[
\begin{array}{cc}
\hat{A}_{i} & 0_{1\times 2}\\
0_{2\times 1} & 0_{2\times 2}
\end{array}
\right]\ ;\quad i=\al,z\equiv\al,3\ ;\quad\al=x,y\equiv1,2 \, ,
\ee
The electric component $A_0$ of the $SU(3)$ connection corresponding to the spherically symmetric Ansatz Eq.~\re{a0} and the Higgs field $\F$ corresponding to Eq.~\re{h}, likewise
\be
\label{axa0}
A_{0}=\left[
\begin{array}{cc}
\hat{A}_{0} & 0_{1\times 2}\\
0_{2\times 1} & 0_{2\times 2}
\end{array}
\right]+i\,v(\rho,z)\,\la_8 \, ,
\ee
and
\be
\label{axh}
(2\eta)^{-2}\,\F=\left[
\begin{array}{cc}
\hat{\F} & 0_{1\times 2}\\
0_{2\times 1} & 0_{2\times 2}
\end{array}
\right]+i\,g(\rho,z)\,\la_8 \, ,
\ee
respectively.

There now remains to impose axial symmetry on the $SU(2)$ algebra valued quantities $\hat{A}_{i}=(\hat{A}_{\al},\hat{A}_{z})$, $\hat{A}_{0}$ and $\hat{\F}$.
For this, we employ the chiral $SO(4)$ matrices~\footnote{\bea
\label{sipm}
\Si_{MN}^{(+)}&=&-\frac14\,(\Si_{M}\tilde\Si_{N}-\Si_{N}\tilde\Si_{M})\label{si+} \, ,\\
\Si_{MN}^{(-)}&=&-\frac14\,(\tilde\Si_{M}\Si_{N}-\tilde\Si_{N}\Si_{M})\label{si-} \, ,
\eea
where the index $M=\al,3,4$, with $\al=1,2$. The spin matrices used are
\be
\label{spin}
\Si_{\al}=-\tilde\Si_{\al}=i\,\si_{\al}\ ,\quad\Si_{3}=-\tilde\Si_{3}=i\,\si_{3}\ ,\quad\Si_{4}=\tilde\Si_{4}=\eins\,,
\ee
where $(\si_{\al},\si_{3})$ are the usual $2\times 2$ Pauli spin matrices.

The matrices $\Si_{MN}^{(\pm)}$ are (anti)self-dual
\be
\label{asd}
\Si_{MN}^{(\pm)}=\pm\,\frac12\,\vep_{MNRS}\,\Si_{RS}^{(\pm)}\,,
\ee
In particular, we opt for the selfdual case.} $\Si_{MN}^{(\pm)}$ representing the $SU_{(\pm)}(2)$ subalgebra valued quantities in Eqs.~\re{axai}-\re{axh}

In this notation,
\bea
\hat A_{\al}&=&\left(\frac{\xi^2+n}{\rho}\right)\,(\vep\hat x)_{\al}\, \Si_{12}+\left[\left(\frac{\xi^1}{\rho}\right)(\vep\hat x)_{\al}(\vep n)_{\ga}
+a_{\rho}\,\hat x_{\al} n_{\ga}\right]\Si_{\ga 3}\label{Aal} \, ,\\
\hat A_{z}&=&a_z\,n_{\ga}\,\Si_{\ga 3}\label{Az}\, ,\\
\hat A_{0}&=&-\chi^1\,n_{\ga}\,\Si_{\ga 4}+\chi^2\,\Si_{34}=\chi^1\,(\vep n)_{\ga}\,\Si_{\ga 3}+\chi^2\,\Si_{12}\label{A0}\, ,\\
\hat \F&=&-\f^1\,n_{\ga}\,\Si_{\ga 4}+\f^2\,\Si_{34}=\f^1\,(\vep n)_{\ga}\,\Si_{\ga 3}+\f^2\,\Si_{12}\label{F}\, ,
\eea
where $n_{\al}=(\cos n\vf,\sin n\vf)$ is the unit vector in the $(x_1,x_2)$ plane, $\vf$ is the azimuthal angle and $n$ is the vortex number.
The functions $(a_{\rho},a_{z})$, $\xi^a=(\xi^1,\xi^2)$, $\chi^a=(\chi^1,\chi^2)$ and $\f^a=(\f^1,\f^2)$ all depend on
the two variables $\rho=\sqrt{|x_{\al}|^2}$ and $z$, and are independant of the time coordinate $x_0$.

The gauge covariant quantities $F_{\mu\nu}=(F_{\al\beta},F_{\al z},F_{\al 0},F_{z 0})$ and $D_{\mu}\F=(D_{\al}\F,D_{z}\F,D_{0}\F)$ follow,
\bea
\hat F_{\al\beta}&=&-\frac{1}{\rho}\,\vep_{\al\beta}\left[D_{\rho}\xi^1\,(\vep n)_{\ga}\,\Si_{\ga 3}+D_{\rho}\xi^2\,\Si_{12}\right]\label{falbeta}\, ,\\
\hat F_{\al z}&=&f_{\rho z}\,x_{\al}n_{\ga}\Si_{\ga 3}-
\frac{1}{\rho}(\vep\hat x)_{\al}\left[D_{z}\xi^1\,(\vep n)_{\ga}\,\Si_{\ga 3}+D_{z}\xi^2\,\Si_{12}\right]\label{faz}\, ,\\
\hat F_{\al 0}&=&\frac{1}{\rho}(\chi\vep\xi)(\vep\hat x)_{\al}n_{\ga}\Si_{\ga 3}+\hat x_{\al}
\left[D_{\rho}\chi^1(\vep n)_{\ga}\Si_{\ga 3}+D_{\rho}\chi^2\,\Si_{12}\right]\label{fal0}\, ,\\
\hat F_{z0}&=&D_{z}\chi^1(\vep n)_{\ga}\Si_{\ga 3}+D_{z}\chi^2\,\Si_{12}\label{z0}\, ,
\eea
and
\bea
D_{\al}\hat \F&=&\hat x_{\al}\left[D_{\rho}\f^1\,(\vep n)_{\ga}\,\Si_{\ga 3}+D_{\rho}\f^2\,\Si_{12}\right]
+\frac{1}{\rho}(\f\vep\xi)(\vep\hat x)_{\al}n_{\ga}\Si_{\ga 3}\label{DFal}\, ,\\
D_{z}\hat \F&=&D_{z}\f^1\,(\vep n)_{\ga}\,\Si_{\ga 3}+D_{z}\f^2\,\Si_{12}\label{DFz}\, ,\\
D_{0}\hat \F&=&(\f\vep\chi)\,n_{\ga}\Si_{\ga 3}\label{DF0}\, ,
\eea
which are all expressed in terms of the $SO(2)$ curvature $$f_{\rho z}=\pa_{\rho}\,a_z-\pa_z\,a_{\rho}\ ,$$ the $SO(2)$ covariant derivatives
\bea
D_{\rho}\xi^a&=&\pa_{\rho}\xi^a+a_{\rho}(\vep\xi)^a\ ,\quad D_{z}\xi^a=\pa_{z}\xi^a+a_{z}(\vep\xi)^a\ ,\quad etc.\nonumber
\eea
and with
\[
(f\vep g)=\vep^{ab}f^ag^b.
\]
The (static) axially symmetric $U(1)\simeq SO(2)$ gauge connection $a_{\mu}=(a_{\al},a_{z},a_{0})$ can be expressed as
\bea
a_{\al}&=&u(r,\ta)\,(\hat{x}\vep)_{\al}\label{aal}\, ,\\
a_{z}&=&0\label{az1}\, ,\\
a_{0}&=&a_{0}(r,\ta)\label{a01} \, .
\eea
In the calcualtion of the angular momentum, the azimuthal component of the Abelian connection $a_{\vf}$ will be employed, which in the notation of Eq.~\re{aal} is
\be
\label{avf}
a_{\vf}=\rho\,u\,.
\ee
The components of the Abelian curvature $h_{\mu\nu}=\pa_{\mu}a_{\nu}-\pa_{\nu}a_{\mu}$ follow
\bea
h_{\al\beta}&=&\frac{1}{\rho}\,(\rho\,u)_{,\rho}\,\vep_{\al\beta}\label{galbeta}\, ,\\
h_{\al z}&=&u_{,z}\,(\vep\hat{x})_{\al}\label{galz}\, ,\\
h_{\al 0}&=&(a_{0})_{,\rho}\,\hat{x}_{\al}\label{gal0}\, ,\\
h_{z 0}&=&(a_{0})_{,z}\label{gz0}\, .
\eea
The reduced two dimensional Lagrangian is
\bea
\label{redHCS1}
L^{(1)}&=&-\frac14\,\left\{\frac{1}{\rho}\left(|D_{\rho}\xi|^2+|D_{z}\xi|^2\right)-\rho
\left(|D_{\rho}\chi|^2+|D_{z}\chi|^2\right)+\rho\,f_{\rho z}^2-\frac{1}{\rho}(\chi\vep\xi)^2
-4\rho\,\left[\pa_{\rho}v^2+\pa_{z}v^2\right]\right\}\nonumber\\
&&\qquad\qquad-\eta^2\left\{\rho\left(|D_{\rho}\f|^2+|D_{z}\f|^2\right)+\frac{1}{\rho}(\f\vep\xi)^2-\rho(\f\vep\chi)^2
-4\rho\,\left[\pa_{\rho}g^2+\pa_{z}g^2\right]\right\}+\ka_1\,\omega^{(1)} \, ,
\eea
where $\omega^{(1)}$ is the reduced two dimensional HCS density Eq.~\re{CS1},
\be
\label{redCS1_ang}
\omega^{(1)}=\frac{8}{\sqrt{3}}\,\eta\left\{g\left[(\chi\vep\xi)\,f_{\rho z}-D_{[\rho}\xi\cdot D_{z]}\chi\right]+
\pa_{[\rho}v(\f\cdot D_{z]}\xi)\right\} \, .
\ee

\begin{small}

\end{small}


\begin{thebibliography}{99}
\bibitem{Navarro-Lerida:2013pua}
  F.~Navarro-L\'erida, E. Radu, and D.~H.~Tchrakian,
  Int. J. Mod. Phys. A {\bf 29} (2014) 1450149
  [arXiv:1311.3950 [hep-th]].
\bibitem{Tchrakian:2010ar}
 D.~H.~Tchrakian,
  J.\ Phys.\ A {\bf 44} (2011) 343001
  [arXiv:1009.3790 [hep-th]].
\bibitem{Radu:2011zy}
  E.~Radu and T.~Tchrakian,
  arXiv:1101.5068 [hep-th].
\bibitem{Jackiw:1985}
see for example, R.~Jackiw, "Chern-Simons terms and cocycles in physics and mathematics",
in E.S. Fradkin $Festschrift$, Adam Hilger, Bristol (1985).
\bibitem{Deser:1982vy}
  S.~Deser, R.~Jackiw and S.~Templeton,
  Phys.\ Rev.\ Lett.\  {\bf 48} (1982) 975.
\bibitem{Paul:1986ix}
  S.~K.~Paul and A.~Khare,
  Phys.\ Lett.\ B {\bf 174} (1986) 420
   [Erratum-ibid.\  {\bf 177B} (1986) 453].
\bibitem{Hong:1990yh}
  J.~Hong, Y.~Kim and P.~Y.~Pac,
  Phys.\ Rev.\ Lett.\  {\bf 64} (1990) 2230.
\bibitem{Jackiw:1990aw}
  R.~Jackiw and E.~J.~Weinberg,
  Phys.\ Rev.\ Lett.\  {\bf 64} (1990) 2234.
\bibitem{Julia:1975ff}
B.~Julia and A.~Zee,
  Phys.\ Rev.\ D {\bf 11} (1975) 2227.
\bibitem{Peccei:1977ur}
  R.~D.~Peccei and H.~R.~Quinn,
  Phys.\ Rev.\ D {\bf 16} (1977) 1791.
\bibitem{Peccei:1977hh}
  R.~D.~Peccei and H.~R.~Quinn,
  Phys.\ Rev.\ Lett.\  {\bf 38} (1977) 1440.

\bibitem{COLSYS}
 U. Ascher, J. Christiansen, R.~D. Russell,
 Mathematics of Computation {\bf 33} (1979) 659;
 ACM Transactions {\bf 7} (1981) 209.
\bibitem{Corrigan:1975hd}
  E.~Corrigan, D.~I.~Olive, D.~B.~Fairlie and J.~Nuyts,
  Nucl.\ Phys.\ B {\bf 106} (1976) 475.
\bibitem{Sinha:1976bw}
  A.~Sinha,
  Phys.\ Rev.\ D {\bf 14} (1976) 2016.
\bibitem{Weinberg}
E.~J.~Weinberg, ``Classical Solutions in Quantum Field Theory: Solitons and Instantons in High Energy Physics'', Cambridge Monographs on Mathematical Physics, Cambridge (2012).
\bibitem{Navarro-Lerida:2014bja}
  F.~Navarro-L\'erida, E.~Radu and D.~H.~Tchrakian, Phys. Rev. D {\bf 90} (2014) 064023.
\bibitem{Skyrme:1962vh}
  T.~H.~R.~Skyrme,
  Nucl.\ Phys.\  {\bf 31} (1962) 556.
 

  

\end{thebibliography}
\end{document}